# Modified Hamiltonian-driven shortest-path ray tracing in dissipative anisotropic media

By

Muhammad F. T. Darwis[1], Fateh Bouchaala[1*], Umair Bin Waheed[2], Mohamed Kamel Riahi[3]

*1. Department of Earth Sciences, Khalifa University of Science and Technology, Abu Dhabi, United Arab Emirates*

*2. Geosciences Department, King Fahd University of Petroleum and Minerals (KFUPM), Dhahran, Saudi Arabia*

*3. Department of Mathematics, Khalifa University of Science and Technology, Abu Dhabi, United Arab Emirates*

*** Corresponding author**: Fateh Bouchaala, email: fateh.bouchaala@ku.ac.ae. ORCID: 0000-0003-2368-3302.

Keywords: Numerical modelling; Body waves; Computational Seismology; Seismic anisotropy; Seismic attenuation; Wave propagation

ABSTRACT

Enhancing ray tracing in dissipative anisotropic media is challenging because the subsurface exhibits both geologic and rheological discontinuities and directionally dependent attenuation. Conventional real-ray tracing and complex energy-velocity formulations struggle with qSV cusps, wavefront triplications, and abrupt discontinuities. Recent conjugate and shortest-path extensions have been proposed to mitigate these issues, yet they rely on approximate Hamiltonian formulations that are typically tested in smooth background models. In addition, they rarely address first-arrival refraction or the coupled behavior of wave propagation and energy dissipation. Here, the g*-Hamiltonian-based formulation is incorporated into a shortest-path-based ray-tracing algorithm for layered dissipative anisotropic media. At the local scale, we first compute the complex energy velocities and corresponding ray quantities, which are ray velocity, ray attenuation, and ray quality factor, using three kernels. These kernels produce complex energy velocities for all body waves in different fashions, namely, the original complex energy velocity (OEV), conjugate real ray tracing (C-RRT), and g*-Hamiltonian (MEV). We then analyze their discrepancies and numerical behavior and examine several optimization strategies. The results demonstrate that all kernels agree for qSH and qP waves, whereas OEV generates pseudo-reflection or numerical instability and unphysical attenuation along ray paths in qSV modeling. C-RRT produces large errors, including non-physical negative attenuation in some qP and qSV traveltime values, and strongly biased ray quality factors under our parameter regime. Among the tested kernels, MEV-driven shortest-path ray tracing produces the most computationally stable first-arrival behavior. These outcomes provide numerical evidence that incorporating the g*-Hamiltonian ray quantities into an MSPM-type framework is a promising technique for first-arrival ray tracing in dissipative anisotropic media.

## INTRODUCTION

Powered by the high-frequency asymptotic approximation, ray tracing techniques are crucial tools for simultaneously modeling the traveltimes of distinct seismic body waves and their raypaths (Cao and Greenhalgh, 1993; Cerveny, 2001a; Huang and Greenhalgh, 2019). A key strength of ray theory in seismic propagation modeling is its ability to capture fundamental seismic wave-propagation mechanisms, such as refraction, transmission, reflection, and mode conversions (Hanyga and Seredyńska, 2000). It also provides physical insights into wave-propagation mechanisms, as the wavefield is decomposed into body waves, enabling their precise identification (Červený and Brown, 2011). In addition, it allows the energy trajectory to be tracked through the medium, which is a pivotal aspect for further analysis in seismic tomography and migration (Pšenčík, Růžek, and Jílek, 2020).

The subsurface is commonly approximated as either elastic (with or without anisotropy) or acoustic, assuming no energy dissipation. Ray-tracing results are widely used in seismic tomography and migration to image the Earth's interior (Gray and May, 1994; Huang and Bellefleur, 2012; Pšenčík, Růžek, and Jílek, 2020). However, rock formations are rarely purely elastic or acoustic and often exhibit anisotropy (Crampin, 1984; Thomsen, 1986; Carcione, 2022). In realistic conditions, anisotropy and attenuation must then be treated jointly and included together in the ray-tracing formulation.

The significance and benefits of implementing attenuation and anisotropy into the ray tracing formulation have been extensively discussed in several publications (Hearn and Krebes, 1990a, 1990b; Chapman et al., 1999; Hanyga and Seredyńska, 2000). Seismic ray tracing in dissipative media was initially formulated within complex ray theory, in which the equations of motion are solved in a complex-valued medium (Hearn and Krebes, 1990a, 1990b; Chapman et al., 1999; Huang et al., 2018). However, this approach can lead to spurious computations and has therefore not been widely implemented in practice (Li et al., 2020). In addition, conditioning the medium to be as complex as this is neither realistic nor physically meaningful. Hence, a robust solution is the real ray tracing (RRT) method (Vavryčuk, 2007b, 2008a, 2012), which solves ray tracing in real space while accounting for attenuation. Thus, the computational complexities caused by the complex space can be alleviated. However, this method rarely provides reliable qSV waves when cusps and triplications occur on the wavefronts in vertical transversely isotropic (VTI) and orthorhombic media (Li et al., 2020a; Wu, Zhou, Bouzidi, et al., 2021; Zhou et al., 2024). To overcome this challenge, notable enhancements to real-space ray tracing (RRT) have been observed, such as the conjugate RRT (C-RRT). The C-RRT employs the first implementation of complex-conjugate eigenvectors for each body wave in the complex energy formulation (Červený and Pšenčík, 2005a, 2005b; Li et al., 2020; Wu et al., 2021a). Hence, these studies claimed to successfully address cusps or triplications in the qSV wavefront, thereby capturing a smooth wavefront in both the propagation and attenuation domains of ray-quantity modeling. However, the formulations in those studies remain approximate and require further analytical justification and improvement. The more rigorous formulation of complex energy velocity, namely the g*Hamiltonian (g*H), which accounts for the asymmetry in the complex density-normalized modulus, has been examined and compared across various lithologies and spatial dimensions. Across these scenarios, the g*H formulation is capable of modeling wavefront and ray quantities, especially the qSV cusp. In addition, these studies strengthen the g*Hamiltonian computation by using optimization functions (i.e., MATLAB routines) to address the complex slowness direction, even though there is no in-depth analysis of the most suitable optimization strategies.

All the aforementioned ray tracing approaches stem from the method of characteristics, formulated canonically in terms of the Hamiltonian (Courant and Hilbert, 2008; Vavryčuk, 2008a; Červený and Brown, 2011). The ray-tracing equations could then be expressed as two distinct sets of partial differential equations. One set governs the change in position (ray paths) with travel time and defines the (complex) ray-velocity formulation. The other set depicts the change in slowness with travel time and shows how it varies along the paths. These equations can be solved using fourth-order Runge–Kutta schemes(Cerveny, 2001; Cerveny and Brown, 2001; Vavryčuk, 2001; Vavryčuk, 2012; Riley, Hobson, and Bence, 2019; Wu, Zhou, Bouzidi, et al., 2021).

However, this method is most applicable in a gradually varying model, due to the second part of the equation, the slowness variation, in which a small perturbation is needed to make the raypaths bend. Hence, this approach is highly sensitive to sharp medium contrasts or edges that might occur along rock boundaries and geological structures (faults, fractures, vugs, etc.). As a result, sharp medium contrasts are effectively replaced by steep gradients. This gradient causes the rays to bend continuously and lose their ability to perform sharp changes, such as refraction, which is one of the fundamental mechanisms governing first arrivals. Recent studies have shown that this method can fail, resulting in distorted raypaths, as illustrated by the dotted line in the original Marmousi2 (Liu, 2025). Thus, the model must be smoothed to remove structural sensitivity while remaining absent from refraction. This mathematical condition is not always geologically realistic or representative of absolute subsurface heterogeneity, in which density contrasts, lateral and vertical discontinuities, and variations in grain and pore size occur randomly. In addition, this method suffers from shadow zones, especially for qSV waves, where regions are depleted of energy, producing blind spots, due to the shooting-bending method's behavior, which demands a carefully selected initial angle of either slowness or ray direction.

Several grid-based methods have been developed to overcome the limitations in the ray-bending techniques, such as extending the MSPM to dissipative anisotropic media (DAM) by combining complex ray quantities from C-RRT (Li et al., 2020; Wu et al., 2021a). Although these studies successfully address ray tracing in dissipative VTI and orthorhombic media, they still rely on an approximate Hamiltonian formulation and an iterative search, and their examples emphasize the use of smooth vertical gradient velocity models. Consequently, they do not clearly discuss the first-arrival refraction events in the heterogeneous viscoelastic anisotropic medium. Hence, what remains unresolved is a ray tracing algorithm that can concurrently account for anisotropy, attenuation, geological heterogeneities, and abrupt discontinuities while using a local ray-quantity formulation that remains stable and robust in the presence of qSV cusps and triplications.

In this study, we advance a MSPM-type ray-tracing framework for layered (piecewise heterogeneous), viscoelastic, anisotropic media by leveraging the g*-Hamiltonian formulation at the point scale for complex-ray computation to construct a field-scale MSPM-type framework. This integration is designed to preserve accurate ray quantities that capture all possible first-arrival events for VTI angles during ray propagation and attenuation, while leveraging a Dijkstra-like modified shortest-path (MSPM-type) algorithm at the field scale to model traveltime fields and backtrace raypaths in a discretized grid network. The yielding framework is tailored for a dissipative VTI medium and is assessed with respect to both local ray-quantity performance and field-scale first-arrival body-wave propagation.

Despite several existing MSPM implementations and ray quantity formulation having been developed (Zhou and Greenhalgh, 2005, 2006; Bai et al., 2007; Zhang and Zhou, 2018; Li et al., 2020), to our limited knowledge, comparative studies of how different local kernels behave in the MSPM or any Dijkstra-like ray tracing remain rare. Hence, to investigate how the selection of local kernel influences complex first-arrival traveltime and reconstructed raypaths, we benchmark the g*-Hamiltonian performance against two closely related alternative kernels: the original complex energy velocity (g-Hamiltonian) and conjugate real-ray tracing (C-RRT) computations at both node and field scales for all body wave modes. The benchmark is limited to these kernels as they are the tightest competitors in constructing energy velocities. In this study, the primary criterion is the physical reasonableness of first-arrival refraction behavior, particularly for qSV, while runtime serves as a complementary yet essential parameter.

The contributions of this study are threefold. First, we implement the g*-Hamiltonian energy-velocity computation in an MSPM-type algorithm for DAM. Second, we benchmark the ray tracing implementation driven by OEV, MEV, and C-RRT kernels at both node and field scales for qP, qSV, and qSH waves in a dissipative VTI medium. Third, we present numerical evidence that, among the kernels tested, the MEV-based ray tracing produces the most stable first-arrival propagation and attenuation in traveltimes and raypaths in the presence of qSV triplication-caused challenges.

## EQUATION OF MOTION AND EIKONAL FORMULATION

In a smoothly heterogeneous DAM without considering any sources, the equation of motion is given as (Červený and Pšenčík, 2005b; Červený and Brown, 2011),

$$\rho\omega^2\boldsymbol{u}_j + \rho a_{ijkl}\partial_{li}\boldsymbol{u}_k = 0, \tag{1}$$

where $\rho(\mathrm{x})$ is the density, $\omega$ is the angular frequency, $\boldsymbol{u}_j(\mathrm{x},\omega)$ is the displacement vector of $j^{th}$ -component, and $a_{ijkl}(\mathrm{x},\omega)$ is the complex-valued fourth-order dissipative moduli normalized by density. In a linear viscoelastic medium, the governing relationship of stress and strain is a constitutive law with memory. In the time domain, the stress-strain relation could be expressed with the relaxation stiffness tensor $C_{ijkl}(s)$ and first-order derivatives of strain $\dot{\varepsilon}_{kl}(t-s)$

$$\sigma_{ij}(t) = \int_0^\infty C_{ijkl}(s) * \dot{\varepsilon}_{kl}(t-s)ds\,. \tag{2}$$

This relation is Hooke's law extended to viscoelastic reality and consistents to standard linear rheologies, namely the Zener model. In the frequency domain, Equation 2 is transformed and expressed in algebraic form

$$\tilde{\sigma}_{ij}(\omega) = \tilde{C}_{ijkl}(\omega)\tilde{\varepsilon}_{kl}(\omega), \tag{3}$$

where the stiffness tensor is complex valued ($\tilde{C}_{ijkl}(\omega) = i\omega\tilde{c}_{ijkl}(\omega)$). The density-normalized complex modulus in the frequency domain $a_{ijkl}(\omega)$ then writes

$$a_{ijkl}(\omega) = \frac{\tilde{C}_{ijkl}(\omega)}{\rho} = a_{ijkl}^{(R)}(\omega) - ia_{ijkl}^{(I)}(\omega) = a_{ijkl}^{(R)}(\omega)\left(1 - \frac{i}{Q_{ijkl}(\omega)}\right), \tag{4}$$

where $a_{ijkl}^{(R)}(\omega)$ depicts the elastic medium property, contributing to wave propagation, $a_{ijkl}^{(I)}(\omega)$ indicates the constitutive source of energy loss corresponding to dissipation within one oscillation cycle, and $Q_{ijkl}(\mathrm{x}, \omega)$ denotes the quality factor. Equation 4 writes the frequency-domain Hooke's law in a DAM, as the effective constitutive representations used in this study. The sign convention remain consisten with the time-harmonic plane wave of this medium, deduced as a high-frequency harmonic signal and expressed as (Vavryčuk, 2007a, 2007b, 2008a),

$$\boldsymbol{u}_i = \mathbf{A}_i e^{\mathrm{i}\omega(\tau - \mathbf{p}_i \cdot \mathrm{x}_i)} = |\mathbf{A}_i|\mathbf{g}e^{\mathrm{i}\omega(\tau - \mathbf{p}_i \cdot \mathrm{x}_i)}, \tag{5}$$

where $\mathbf{A}_i$, $\boldsymbol{p}_i$, and $\boldsymbol{u}_i$ are the Cartesian components, depicted as the subscript $i$ of the complex ray amplitude vector, complex-valued slowness vector, and displacement vector, respectively**.** Meanwhile, $\mathbf{g}$ is the vector of polarization direction or the complex eigenvector. In addition, let $\tau$ denote the complex traveltime that reads

$$\boldsymbol{\tau}(\mathrm{x}) = \boldsymbol{\tau}^{(R)}(\mathbf{x}) + i\boldsymbol{\tau}^{(I)}(\mathbf{x}), \tag{6}$$

in which the superscripts *R* and *I* depict the real and imaginary parts, respectively, and $i$ is the imaginary unit in this case. The $\boldsymbol{\tau}^{(R)}(\mathbf{x})$ defines the propagation traveltime while $\boldsymbol{\tau}^{(I)}(\mathbf{x})$ defines the attenuation traveltime units (Vavryčuk, 2007b, 2008b, 2008a; Zhou et al., 2024). This representation is favorable as

$$e^{\mathrm{i}\omega\boldsymbol{\tau}(\mathrm{x})} = e^{\mathrm{i}\omega\left(\boldsymbol{\tau}^{(R)}(\mathbf{x})+\, i\boldsymbol{\tau}^{(I)}(\mathbf{x})\right)} = e^{\mathrm{i}\omega\left(\boldsymbol{\tau}^{(R)}(\mathbf{x})\right)} . e^{-\omega\left(\boldsymbol{\tau}^{(I)}(\mathbf{x})\right)}. \tag{7}$$

Hence, equation 7 distinguishes the oscillatory and energy decay parts, where the phase factor $e^{\mathrm{i}\omega\left(\boldsymbol{\tau}^{(R)}(\mathbf{x})\right)}$ denotes the wavefront propagation and therefore corresponds to the propagation traveltime $\boldsymbol{\tau}^{(R)}(\mathbf{x})$. Meanwhile, the factor $e^{\mathrm{i}\omega\left(\boldsymbol{\tau}^{(I)}(\mathbf{x})\right)}$ represents the exponential amplitude decay and thus depicts $\boldsymbol{\tau}^{(I)}(\mathbf{x})$, which is a measure of attenuation expressed in time-like units. Positive values of $\boldsymbol{\tau}^{(I)}(\mathbf{x})$ represent the amplitude decay, and negative values indicate exponential amplification. In this expression, $\boldsymbol{\tau}^{(I)}(\mathbf{x})$ is not a different physical arrival time but an energy-loss representation in time. Rather, it is a quantity in time units, with $\omega$ yielding the attenuation exponent. In other words, the attenuation in traveltime is therefore the ray-based manifestation of the constitutive loss term $a^{(I)}_{ijkl}$, where $a^{(I)}_{ijkl} \geq 0$ results in $\boldsymbol{\tau}^{(I)} \geq 0$. Negative $\boldsymbol{\tau}^{(I)}$ would represent amplitude amplification and is thus considered non-physical when the given $a^{(I)}_{ijkl} \geq 0$.

For the rest of the paper, the notation $\boldsymbol{\tau}^{(R)}(\mathbf{x})$ and $\boldsymbol{\tau}^{(I)}(\mathbf{x})$ are referred to as propagation and attenuation in traveltime, respectively. Plugging equation 5 into equation 1 elicits the Christoffel $\boldsymbol{\Gamma}$ and Eikonal equation

$$\left[\boldsymbol{\Gamma}_{jk}(\mathrm{x},\mathbf{p}) - \delta_{jk}\right]\mathbf{g}_k = 0,\ \ \boldsymbol{\Gamma}_{jk}(\mathrm{x},\mathbf{p}) = a_{ijkl}\mathbf{p}_l\mathbf{p}_i, \tag{8}$$

where $\delta_{jk}$ is the Kronecker delta, and$\mathbf{p}$ is the complex slowness vector. As $\mathbf{p} = \frac{\mathbf{n}}{c}$ (Vavryčuk, 2007b), The Eikonal equation becomes,

$$\left[\boldsymbol{\Gamma}_{jk}(\mathrm{x},\mathbf{n}) - c^2\delta_{jk}\right]\mathbf{g}_k = 0,\ \ \boldsymbol{\Gamma}_{jk}(\mathrm{x},\mathbf{n}) = a_{ijkl}\mathbf{n}_l\mathbf{n}_i, \tag{9}$$

where $\mathbf{n}$ denotes the complex unit slowness direction vector, $c^2$ and $c$ are complex-valued, corresponding to the eigenvalue and phase velocity of a body wave, respectively. A non-trivial polarization vector $\mathbf{g}$ exists only if the homogeneous system in equation 9 admits non-zero solutions, which requires:

$$\det\left[\mathbf{\Gamma}_{jk}(\mathrm{x}, \mathbf{n}) - c^2\delta_{jk}\right] = 0 \tag{10}$$

In a general DAM, it can be expressed in terms of complex angles using sine and cosine functions as (Hanyga and Seredyńska, 2000),

$$\mathbf{n} = \begin{bmatrix} \sin(\theta + i\vartheta)\cos(\varphi + i\psi) \\ \sin(\theta + i\vartheta)\sin(\varphi + i\psi) \\ \cos(\theta + i\vartheta) \end{bmatrix} = \mathbf{n}^{(R)} + i\mathbf{n}^{(I)}, \tag{11}$$

where $(\theta, \vartheta)$ and $(\varphi, \psi)$ are pairs of inclination and azimuthal angle pairs in 3D space. These angle pairs are required to be determined from the known $a_{ijkl}$ and $Q$ factors (Wu et al., 2021b; Zhou et al., 2024). With the known $a_{ijkl}$ obtained, for example, from petrophysical measurements or previous studies, $\mathbf{n}$ could be solved from equation 11 by specifying the directions of $\theta$ and $\varphi$ and performing a trial of $\vartheta$ and $\psi$ to solve equation 10 and obtain the eigenvalues $c^2$ and $\mathbf{p}$. Subsequently, $\mathbf{g}$ could be obtained by solving equation 9. Solving for the eigenvalues and eigenvectors is paramount in this study, as they are used to define the complex energy velocities. In the dissipative VTI case, the azimuthal angles vanish (i.e., $\vartheta = \psi = 0$); thus, equation 11 could be simplified to

$$\mathbf{n} = \begin{bmatrix} \sin(\theta + i\vartheta) \\ 0 \\ \cos(\theta + i\vartheta) \end{bmatrix}, \tag{12}$$

which is applicable both in 3D and 2D dissipative VTI cases.

## COMPLEX ENERGY VELOCITY AND COMPLEX RAY QUANTITIES

In what follows, we refer to $\mathbf{v}^{(R)}$ as the propagation energy velocity, $\mathbf{v}^{(I)}$ as the attenuation in energy velocity units. Stems from these, we denote $V^{ray}$ as the ray velocity describing the propagation velocity along a ray, $A^{ray}$ as the ray attenuation, which defines the weakening of amplitude along a ray, and $Q^{Ray}$ as the ray Q-factor (Vavryčuk, 2007b, 2008b) as the ray quality factor. The last three define the complex ray quantities. Let the body waves propagate in the 3D real-space raypath $\mathbf{x} = (x_1, x_2, x_3)$, with the traveltime definition in equation 6, the velocity in a complex medium is defined as,

$$\mathbf{v} = \frac{d\mathrm{x}}{d\tau} = \frac{d\mathrm{x}}{d\tau^{(R)} + d\tau^{(I)}} = \mathbf{v}^{(R)} + i\mathbf{v}^{(I)} \tag{13}$$

where

$$\mathbf{v}^{(R)} = \frac{1}{1+\alpha^2}\frac{d\mathrm{x}}{d\tau^{(R)}}, \qquad \mathbf{v}^{(I)} = \frac{\alpha}{1+\alpha^2}\frac{d\mathrm{x}}{d\tau^{(I)}} \tag{14}$$

Equation 14 shows that $\mathbf{v}^{(R)} = \alpha\mathbf{v}^{(I)}$, indicating that $\mathbf{v}^{(R)}$and $\mathbf{v}^{(I)}$ are parallel to each other. Since the waves are propagating in the real space, both parts of $\mathbf{v}$ are defined by the same real-valued direction. Therefore, $\mathbf{v}$ is referred to as the homogeneous complex energy velocity vector (Vavryčuk, 2007b), which coincides with the group velocity in elastic media. Let $\hat{\mathbf{r}}^{(R)}$ and $\hat{\mathbf{r}}^{(I)}$ denote the ray direction vectors such that $\mathbf{v}^{(R)} = v^{(R)}\hat{\mathbf{r}}^{(R)}$ and $\mathbf{v}^{(I)} = v^{(I)}\hat{\mathbf{r}}^{(I)}$, by substituting the homogeneous relation between the real and imaginary parts into equation 13, the traveltime perturbation (Zhou et al., 2024) can be expressed as,

$$d\boldsymbol{\tau} = \frac{ds}{\mathbf{v}.\hat{\mathbf{r}}} = \frac{ds}{[v^{(R)}(\hat{\mathbf{r}}^{(R)}.\hat{\mathbf{r}}) + iv^{(I)}(\hat{\mathbf{r}}^{(I)}.\hat{\mathbf{r}})]} = \frac{ds}{V^{Ray}} + iA^{Ray}ds, \tag{15}$$

where $\hat{\mathbf{r}}$ is the complex unit ray direction vector, and

$$V^{Ray} = \frac{\left[v^{(R)}\right]^2 + \left[v^{(I)}\right]^2}{v^{(R)}}, \qquad A^{Ray} = \frac{-v^{(I)}}{[v^{(R)}]^2 + [v^{(I)}]^2}, \qquad Q^{Ray} = -\frac{\mathbf{v}^{(R)}}{\mathbf{v}^{(I)}}. \tag{16}$$

All of these represent ray quantities that characterize how a wave propagates as a ray (Vavryčuk, 2007a, 2007b). Accordingly, after obtaining homogeneous $\mathrm{v}^{(R)}$, and $\mathrm{v}^{(I)}$, equation 16 is applied to determine these ray quantities, which are crucial parameters for performing the traveltime and raypaths computation in DAM (Li et al., 2020; Wu et al., 2021a). Therefore, accurately computing $\mathbf{v}$ is pivotal for initiating ray tracing in VEAM. However, although equation 14 specifies the homogeneity condition in the internal $\mathbf{v}$ that must be honored, it cannot be used to calculate the homogeneous $\mathbf{v}$ due to the unknowns $\frac{d\mathbf{x}}{d\tau}$, and $\alpha$. Consequently, another method must be employed in this case.

## ORIGINAL ENERGY VELOCITY (OEV) VECTOR FORMULATION

Substituting the dot product of $\mathbf{g}$ and the normalization condition $\mathbf{g}.\mathbf{g} = 1$ into equation 9 yields the eikonal equation (Vavryčuk 2008), which can be written as,

$$G(\mathrm{x}, \mathbf{p}) = a_{ijkl}\mathbf{p}_i\mathbf{p}_l\mathbf{g}_j\mathbf{g}_k = 1 \tag{17}$$

Stems from equation 8, the Eikonal equation can be rewritten in the form of a Hamiltonian (Cerveny,2001) as

$$H(\mathrm{x}, \mathbf{p}) = \frac{1}{2}[\mathbf{p}.\Gamma(\mathbf{g})\mathbf{p} - 1] = \frac{1}{2}[G(\mathbf{x}, \mathbf{p}) - 1] = 0, \tag{18}$$

where $H = H(\mathbf{x}, \mathbf{p})$ is referred to as the Hamiltonian, representing the nonlinear partial differential equation for solving the traveltime, and $\mathbf{g}$ is the complex eigenvector. Based on the characteristic method (Courant and Hilbert, 2008), the vector $\mathbf{v}$ can be formulated as (Červený and Brown, 2001; Vavryčuk, 2008b)

$$\mathbf{v} = \frac{d\mathrm{x}_i}{d\tau} = \frac{1}{2}\frac{\partial G}{\partial \mathbf{p}_i} = \frac{\partial H}{\partial \mathbf{p}_i} = a_{ijkl}\mathbf{p}_l\mathbf{g}_j\mathbf{g}_k \tag{19}$$

where $G$ is obtained from equation 17. Equation 19 belongs to the original real-space ray tracing formulation, as used in one of the earliest ray tracing studies applied in VEAM (Vavryčuk, 2007b). Because the vector $g$ appears in the Hamiltonian formulation, previous studies have referred to this approach as the g-Hamiltonian (Zhou et al., 2024). To avoid misrepresenting the original studies by using different terminology, we retain the term "energy velocity" (Vavryčuk, 2007b) and refer to it as the original energy velocity (OEV). Several studies have found limitations in modeling the qSV-wave ray quantities in dissipative transversely isotropic media using OEV (Li et al., 2020; Wu et al., 2021a, 2021b; Zhou et al., 2024). Hence, this method is further examined and compared with our newly developed approach for raypath and traveltime modeling within the ray-tracing framework. In addition, this method does not yield a reliable qP-wave solution for multi-arrival events (transmission, mode conversion, and reflection) at a boundary layer between two VEAMs, due to coupling between P- and qSV-wave propagation (Wu et al., 2022).

## MODIFIED ENERGY VELOCITY (MEV) VECTOR FORMULA

Substituting the dot product of the complex conjugate $\mathbf{g}^*$ together with the normalization condition $\mathbf{g}^*.\mathbf{g} = 1$ into equation 8 yields the eikonal equation, written as

$$\mathbf{p}.\tilde{\mathbf{\Gamma}}\mathbf{p} = 1, \tag{20}$$

where

$$\tilde{\mathbf{\Gamma}}_{il} = a_{ijkl}\mathbf{g}_j\mathbf{g}_k^*, \tag{21}$$

Accordingly, the complex energy velocity can be approximated as

$$\mathbf{v} \approx \tilde{\mathbf{\Gamma}}\mathbf{p} \tag{22}$$

which is referred to as the C-RRT approximation (Wu et al., 2021a, 2021b), and the following identity holds

$$\mathbf{g}^*.\mathbf{\Gamma}(\mathbf{p})\mathbf{g} = \mathbf{g}.\mathbf{\Gamma}(\mathbf{p})\mathbf{g}^* = \mathbf{p}.\tilde{\mathbf{\Gamma}}^{\mathrm{T}}\mathbf{p} = 1. \tag{23}$$

In this equation, the "T" represents the matrix transpose without complex conjugation. However, the matrix $\tilde{\mathbf{\Gamma}}_{il}$ is, in general, asymmetric $\tilde{\mathbf{\Gamma}}_{il} \neq \tilde{\mathbf{\Gamma}}_{li}$. Integrating equations 20 and 21 yields

$$\mathbf{p}.\bar{\bar{\boldsymbol{\Gamma}}}\mathbf{p} = 1, \qquad \bar{\bar{\boldsymbol{\Gamma}}} = \frac{1}{2}\left(\tilde{\boldsymbol{\Gamma}} + \tilde{\boldsymbol{\Gamma}}^T\right). \tag{24}$$

Therefore, equation 24 could be expressed in a new Hamiltonian form:

$$H(\mathrm{x}, \mathbf{p}) = \frac{1}{2}\left(\mathbf{p}.\bar{\bar{\boldsymbol{\Gamma}}}\mathbf{p} - 1\right), \tag{25}$$

which leads to the g*-Hamiltonian definition of complex energy velocity vector, known as g*Hamiltonian or g*H formula, as

$$\mathbf{v} = \frac{\partial H}{\partial \mathbf{p}} = \bar{\bar{\boldsymbol{\Gamma}}}\mathbf{p} + \frac{1}{2}\mathbf{g}^*.\left[\boldsymbol{\Gamma}(\mathbf{p}) - \mathbf{I}\right]\frac{\partial \mathbf{g}}{\partial \mathbf{p}}. \tag{26}$$

Based on extensive numerical investigations using the identity $\mathbf{p}.\mathbf{v} = 1$ and equation 24, the solution becomes unique when $\frac{1}{2}\mathbf{g}^*.\left[\boldsymbol{\Gamma}(\mathbf{p}) - \mathbf{I}\right]\frac{\partial \mathbf{g}}{\partial \mathbf{p}} = \mathbf{0}$ (Zhou et al., 2024). Consequently, the final form complex energy velocity vector is given by as follows,

$$\mathbf{v} = \bar{\bar{\boldsymbol{\Gamma}}}\mathbf{p} \tag{27}$$

Due to the utilization of $\mathbf{g}^*$ in the Hamiltonian configuration, this approach is referred to as g*-Hamiltonian. To maintain consistency with the original definition, in this study, we refer to this method as the modified energy velocity (MEV).

## FERMAT'S VARIATIONAL PRINCIPLE

Fermat's variational principle in VEAM is one of the foundations used in grid-based ray tracing, especially in the MSPM algorithm. This principle has been shown, in general anisotropic media, to be mathematically applicable to the Hamiltonian formulation and variational equations (Bóna and Slawinski, 2003). Based on the formulations of the complex slowness vector ($\mathbf{p} = \nabla\tau$) and the complex energy velocity vector ($\mathbf{v} = \frac{dx}{d\boldsymbol{\tau}}$), the complex traveltime $\boldsymbol{\tau}$ is formulated along the raypath $R$ as,

$$\boldsymbol{\tau} = \int_{\mathrm{R}} \mathbf{p}\, d\mathrm{x} = \int_{\mathrm{R}} \frac{ds}{\mathbf{v}}, \tag{28}$$

where $\mathrm{d}s = |d\boldsymbol{x}|$ is a small segment of the raypath.

In a non-attenuative medium, equation 28 indicates that the slowness vector can be substituted with the energy or group velocity to calculate the traveltime. However, when attenuation is present, both **p** and **v** become complex-valued, yet equation 14 indicates that homogeneity in **v** must always be guaranteed. These conditions give them distinct behaviors.

Among several tests on various fundamental rock properties in dissipative TI media, inhomogeneity in **p** can appear even though **v** is homogeneous (Wang et al., 2024). Thus, we focus on building the traveltime based on **v**. Substituting equations 6 and 15 into 26 yields

$$\boldsymbol{\tau} = \int_{\mathrm{R}} \frac{v^{(R)}}{[v^{(R)}]^2 + [v^{(I)}]^2} \mathrm{d}s + i \int_{\mathrm{R}} \frac{-v^{(I)}}{[v^{(R)}]^2 + [v^{(I)}]^2} \mathrm{d}s = \int_{\mathrm{R}} \frac{1}{V^{Ray}} \mathrm{d}s + i \int_{\mathrm{R}} A^{Ray} \mathrm{d}s, \tag{29}$$

and,

$$\boldsymbol{\tau}^{(R)}(\mathrm{x}) = \int_{\mathrm{R}} \frac{1}{V^{Ray}} ds, \ \boldsymbol{\tau}^{(I)}(\mathrm{x}) = \int_{\mathrm{R}} A^{Ray} ds, \tag{30}$$

Perturbing both sides of equation 29 results in

$$\delta\boldsymbol{\tau}(\mathrm{x}_{AB}) = \delta\left(\int_{\mathrm{A}}^{\mathrm{B}} \frac{1}{V^{Ray}} ds + i \int_{\mathrm{A}}^{\mathrm{B}} A^{Ray} ds\right) = \delta\left(\int_{\mathrm{A}}^{\mathrm{B}} \frac{1}{V^{Ray}} ds\right) + i\delta\left(\int_{\mathrm{A}}^{\mathrm{B}} A^{Ray} ds\right) = 0, \tag{31}$$

where a raypath $R$ originates at point A and terminates at point B in a fixed medium. Equation 31 outlines Fermat's variational principle and infers that a true raypath should be the one that renders the traveltime integer stationary condition $\delta\tau(\mathrm{x}_{AB}) = 0$ (Zhou and Greenhalgh, 2005, 2006). This principle states that the traveltime from a source to a receiver must be minimal (or stationary) among all plausible paths. As a result, taking the traveltime definition from equation 30, the first-arrival propagation traveltime at points A to B is constructed using $V^{Ray}(\mathrm{x}, \hat{\mathbf{r}}_{AB})$ as,

$$\boldsymbol{\tau}_{AB}^{(R)}(\mathrm{x}) = \min\left[\int_{\mathrm{B}}^{\mathrm{A}} \frac{1}{V^{Ray}(\mathrm{x}, \hat{\mathbf{r}}_{AB})} \mathrm{d}s, \forall\, \mathrm{x}_A \in \Omega_B\right], \tag{32}$$

while the attenuation in traveltime is expressed as,

$$\boldsymbol{\tau}^{(I)}(\mathrm{x}) = \min\left[\int_{\mathrm{B}}^{\mathrm{A}} A^{Ray}(\mathrm{x}, \hat{\mathbf{r}}_{AB}) \mathrm{d}s, \forall\, \mathrm{x}_A \in \Omega_B\right], \tag{33}$$

where $\Omega_B$ is in the vicinity of $\mathrm{x}_A$, and the attenuation functional is the imaginary part of the complex traveltime, in line with equation 6.

Note that $V^{Ray}$ and $A^{Ray}$ are now functions of space and complex-valued ray direction $\hat{\mathbf{r}}_{AB}$ reflecting the dissipative and anisotropic conditions of the medium. Obviously, in isotropic cases, $V^{Ray}$ and $A^{Ray}$ can be taken as independent of the ray direction $\hat{\mathbf{r}}_{AB}$. Therefore, the subsequent raypath could be specified as

$$\mathcal{R}_{AB}^{(p)} = \arg \min_{\mathbf{x}_A \in \Omega_B} \left\{\int_{\mathrm{B}}^{\mathrm{A}} \frac{1}{V^{Ray}(\mathrm{x}, \hat{\mathbf{r}}_{AB})} ds\right\}, \; \mathcal{R}_{AB}^{(a)} = \arg \min_{\mathbf{x}_A \in \Omega_B} \left\{\int_{\mathrm{B}}^{\mathrm{A}} A^{Ray}(\mathrm{x}, \hat{\mathbf{r}}_{AB}) ds\right\}, \tag{34}$$

where $\mathcal{R}_{AB}^{(p)}$ is the propagation raypath, describing trajectories along which energy is propagated, and the traveltime is minimum (stationary), while $\mathcal{R}_{AB}^{(a)}$ is the attenuation raypath, representing trajectories along which signal damping is maximum (stationary). Equations 32–34 outline the ray-tracing formulation for any combination of anisotropy and dissipative properties. Nonetheless, a practical technique is required to comprehensively obtain minimum traveltimes and determine first-arrival raypaths between prescribed endpoints.

Once the path integrals are determined on a graph of reasonable piecewise-linear components, Fermat's variational formulation is discretized based on the shortest-path algorithm applicable in an anisotropic medium (Zhou and Greenhalgh, 2005). When the medium is discretized into nodes and cells, according to equations 32 and 33, the minimum traveltime from point $A$ to $B$ in a heterogeneous dissipative VTI medium may be approximated as

$$\begin{gathered}\boldsymbol{\tau}_B^{(R)} \approx \min_{\mathbf{x}_A \in \Omega_B} \left\{ \boldsymbol{\tau}_A^{(R)} + \frac{|\mathrm{x_A} - \mathrm{x_B}|}{V^{Ray}(A, \hat{\mathbf{r}}_{AB}) + V^{Ray}(\mathrm{x_B}, \hat{\mathbf{r}}_{AB})} \right\} \\ \boldsymbol{\tau}_B^{(I)} \approx \min_{\mathbf{x}_A \in \Omega_B} \left\{ \boldsymbol{\tau}_A^{(I)} + |\mathrm{x_A} - \mathrm{x_B}| \left( A^{Ray}(A, \hat{\mathbf{r}}_{AB}) + A^{Ray}(\mathrm{x_B}, \hat{\mathbf{r}}_{AB}) \right) \right\}\end{gathered} \quad (35)$$

where $\Omega_B$ is a group of the neighboring nodes of the unspecified node $B$. Equation 33 is the discrete minimization of the continuous Fermat minimization (Equations 30 and 31).

## RAY TRACING METHOD BASED ON MSPM

The proposed algorithm is a grid-based ray tracing scheme computationally inspired by the shortest-path method and the modified shortest-path method (MSPM) for determining raypaths and wavefronts in gridded media (Zhou and Greenhalgh, 2005; Bai, Greenhalgh, and Zhou, 2007). The medium is defined by the complex density-normalized modulus on a set of grid networks. Fermat's principle serves as the foundation and yields the variational condition of the true raypath as described in the previous section.

In addition, the MSPM also leverages Huygens' principle (Musgrave, 1970), which characterizes wave propagation by supposing that each point on a wavefront serves as an elementary source of secondary wavelets, and the resulting wavefront is determined by the superposition of effects from all the points on the previous wavefront. This principle is then combined with Fermat's principle in equations 32 and 33, indicating that the first arrival travel time at a location in a DAM is the smallest traveltime among all adjacent points. Based on these foundations, we develop the ray tracing algorithm summarized in the following procedures.

## Model Discretization and Traveltime Initialization

The 2D DAM is discretized into a mesh-based model constructed with a number of non-overlapping blocks $N_b$ with specified dimensions ($\mathrm{d}x$ and $\mathrm{d}z$). Each block incorporates an encoded number of nodes $N_n$ along its edges. We implement index notation for $N_b$ blocks by $j \in \mathcal{B} = \{1,2,\ldots,N_b\}$, and for $N_n$ grid nodes by $k \in \mathcal{N} = \{1,2,..,N_n\}$ defining the spatial coordinate $\mathrm{x}_k = (x_k, z_k)$. Each block $j$ is linked to an interior depth and is therefore allocated to the $l^{th}$ $-$layer, facies, or rock type. To construct the non-overlapping condition, shared nodes between neighboring cells are indexed through a coordinate map. Therefore, the node indices are unique and correspond to unique geometric coordinates $\mathrm{x}_j = (x_j, z_j)$.

We denote generic node indices by $i, j$, and source node indices by $i_s$, and receiver indices by $i_r$, to distinguish between their associated block indices and node indices. For each source location $\mathrm{x}_s = (x_s, z_s)$, the framework computes the first-arrival complex traveltime and tracebacks the corresponding raypahts from all receiver locations $\mathrm{x}_r = (x_r, z_r)$ to a source point.

Based on the block and node IDs, at node $j$ from source $s$, the complex traveltime of wave mode $m$ is denoted as

$$\tau_j^{(s,m)} = \tau_j^{(s,m,R)} + i\tau_j^{(s,m,I)}, \tag{36}$$

while, at block $k$ and source $s$, it is defined as

$$T_k^{(s,m)} = T_k^{(s,m,R)} + iT_k^{(s,m,I)}, \tag{37}$$

where each part corresponds to the minimum node traveltime within a block,

$$T_k^{(s,m,R)} = \min_{j\in\mathcal{N}(k)} \tau_j^{(s,m,R)}, \quad T_k^{(s,m,I)} = \min_{j\in\mathcal{N}(k)} \tau_j^{(s,m,I)}. \tag{38}$$

Both $T_k^{(s,m,R)}$, and $T_k^{(s,m,I)}$ independently drive the computation to determine the following block among the neighboring blocks using a Dijkstra-like algorithm that expands by selecting the block with the smallest current traveltime. Furthermore, the corresponding propagation raypaths from the source to the receiver of the wave mode $m$ are represented as a sequence of node indices

$$\mathcal{R}_{s,r}^{(s,m,p)} = \left(i_1^{(s,m,p)}, i_2^{(s,m,p)}, \dots, i_{D^{(m,p)}}^{(s,m,p)}\right), \quad (39)$$

where $i_1^{(s,m,p)} = i_s$, $i_{D^{(m,p)}}^{(s,m,p)} = i_r$ and $D^{(m,p)}$ is the total number of nodes traversed by the propagation raypath of the wave mode $m$.

Equivalently, the attenuation raypath is defined as,

$$\mathcal{R}_{s,r}^{(s,m,a)} = \left(i_1^{(s,m,a)}, i_2^{(s,m,a)}, \dots, i_{D^{(m,a)}}^{(s,m,a)}\right), \quad (40)$$

where $i_1^{(s,m,a)} = i_s$ and $i_{D^{(m,a)}}^{(s,m,a)} = i_r$.

Therefore, the minimum propagation and attenuation in traveltime are computed at each node together with the predecessor node that produced the minimum traveltime**.** Here, following the MSPM philosophy, nodes are classified into primary nodes at the block corners and secondary nodes, spaced along the block edges. The primary nodes calculate energy velocities for all possible take-off angles, corresponding to the nodes in the corners where medium parameters (i.e., petrophysical properties, attenuation, etc.) are sampled directly. Meanwhile, any additional nodes along the edges constructed between the primary nodes are assigned to the secondary nodes to enhance angular coverage, possibilities, and resolution without refining the grid dimensions. The cell parameterization is done by interpolating these primary nodes. In other words, the primary nodes determine the model parameterization, whereas the secondary nodes improve the resolution of the raypath discretization.

Here, we construct a gridded ray-tracing model by embedding each node with complex elastic moduli and density $\left(c_{ijkl}(l),\, Q_{ijkl}(l), \rho(l)\right)$ according to their $l$ th rock type. This configuration facilitates controlling the complexity of discretizing the model and the assumed medium, including the dissipative anisotropic parameters (Fig. 1). If the medium is isotropic or visco-isotropic, the number of independent moduli is reduced. Hence, the velocity formulation in terms of complex group velocities of P- and S-waves is simplified, whereas in a non-attenuating medium, the quality factor term is omitted.

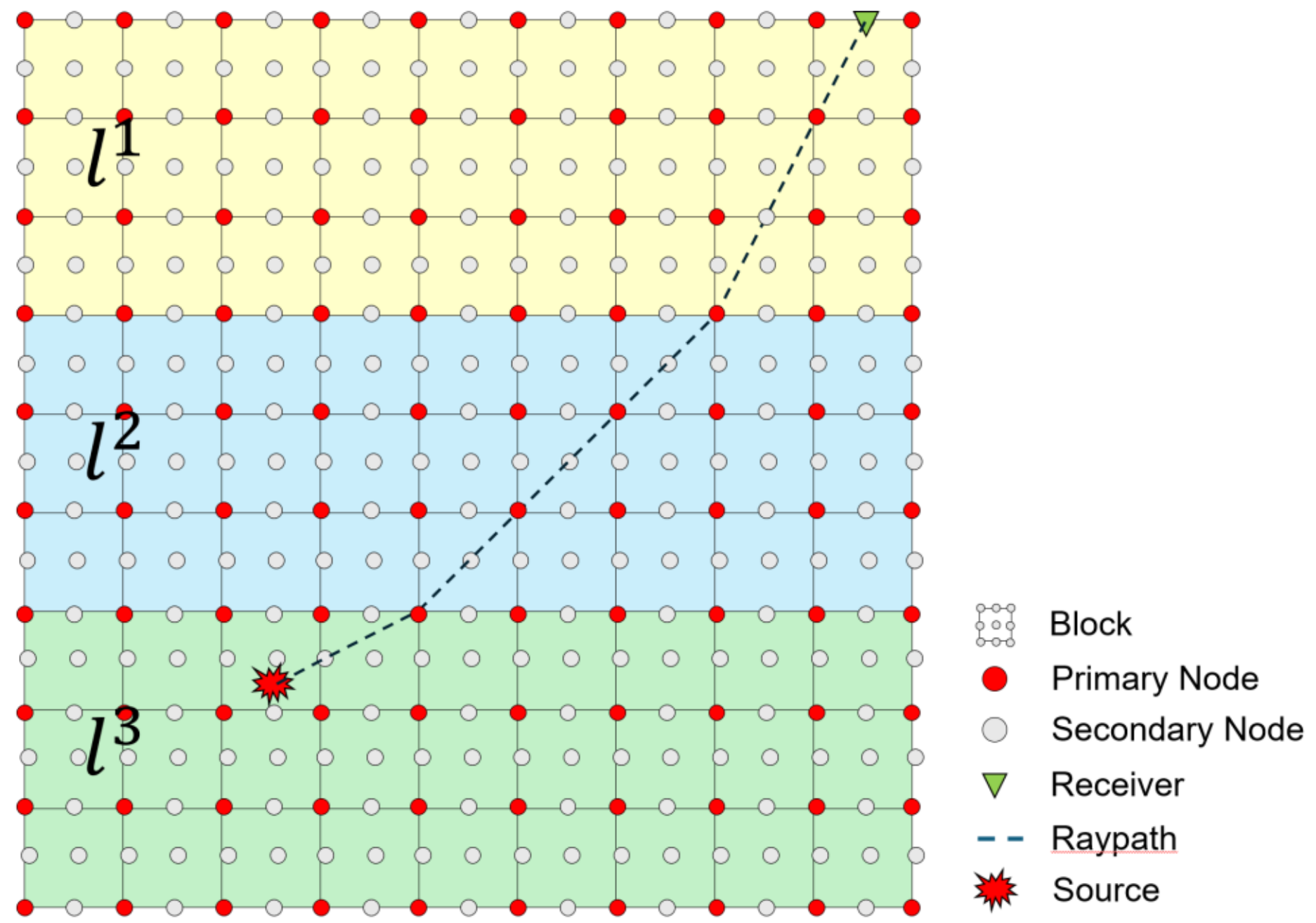

Fig. 1. Example of the block-and-node model discretization used by the MSPM ray tracing. Red and gray dots denote primary and secondary nodes, respectively, and the dashed line illustrates a candidate raypath connecting a source and receiver. The figure shows how local ray-property tables are attached to nodes and how the continuous medium is converted into a graph for minimum-traveltime propagation.

### Minimum Traveltime Computation

The traveltime computation leverages two key components. The first component is a block-wise marching scheme that iteratively identifies the unvisited block with the smallest comparative traveltime. The second component is a local relaxation operator that enforces Fermat's principle within that block by updating the internal nodes. The subsequent raypath is reconstructed by backtracking along the recorded predecessor nodes, which store the neighboring node ID with the smallest traveltime update.

At each iteration, after traveltime values for nodes and blocks have been initialized, the block $k$ is selected among the unvisited blocks with the smallest block traveltime, which initially corresponds to the source block $T_{k_s}^{(s,m,\odot)}$. This source block is then "frozen," and its nodes are relaxed. Subsequently, the neighboring blocks are identified, and their comparative travel times are updated based on the recently updated node values. Inside the current block, i.e, the source block $k_s$, for each arbitrary pair of nodes $i, j$ in block $k$, the distance $\mathbf{d}_{ij}$ between them as assumed to be Euclidean distance, and a raypath candidate is defined as,

$$\mathbf{d}_{ij} = \hat{\mathbf{d}}_{ij}.\mathrm{d}_{ij} = \hat{\mathbf{d}}_{ij}\left|\left|\mathrm{x}_j - \mathrm{x}_i\right|\right|, \qquad \hat{\mathbf{d}}_{ij} = (\mathrm{x}_j - \mathrm{x}_i)/\left|\left|\mathrm{x}_j - \mathrm{x}_i\right|\right|, \tag{41}$$

where the direction $\hat{\mathbf{d}}_{ij}$ determines the raypath unit vector direction within the grid.

For the current wave mode $m$ computation, the algorithm retrieves the ray-property tables, obtained from the OEV and MEV kernel computations, to generate the ray quantities and directions $\hat{\mathbf{r}}_i(m,\theta)$, $\hat{\mathbf{r}}_j(m,\theta)$, $V_i^{Ray}(m,\theta)$, $V_j^{Ray}(m,\theta)$, $A_i^{Ray}(m,\theta)$ and $A_j^{Ray}(m,\theta)$ tabulated over take-off angle $\theta$. The first two quantities are matched with $\hat{\mathbf{d}}_{ij}$ to find the corresponding take-off angles $\theta_i$and $\theta_j$ from precomputed tabulated data, which is crucial to choose for selecting the associated $V_i^{Ray}(m,\theta_i)$, $V_j^{Ray}(m,\theta_j)$, $A_i^{Ray}(m,\theta_i)$ and $A_j^{Ray}(m,\theta_j)$. Furthermore, the segment-averaged ray quantities are computed as,

$$\begin{aligned}\bar{V}^{Ray}(m,\hat{\mathbf{d}}_{ij}) &= \frac{1}{2}\left(V_i^{Ray}(m,\theta_i) + V_j^{Ray}(m,\theta_j)\right),\\ \bar{A}^{Ray}(m,\hat{\mathbf{d}}_{ij}) &= \frac{1}{2}\left(A_i^{Ray}(m,\theta_i) \text{ and } A_j^{Ray}(m,\theta_j)\right),\end{aligned} \tag{42}$$

where the averaged ray velocity $\bar{V}^{Ray}(m, \hat{\mathbf{d}}_{ij})$ and ray attenuation $A^{Ray}(m, \hat{\mathbf{d}}_{ij})$ are utilized to determine the traveltime increment along $\mathbf{d}_{ij}$ in which, for the propagation parameter, this increment takes the form,

$$\Delta\tau_{i\leftrightarrow j}^{(s,m,p)} = \frac{\left\|\mathrm{x}_j - \mathrm{x}_i\right\|}{\bar{V}^{Ray}(m, \hat{\mathbf{d}}_{ij})}. \tag{43}$$

For the attenuation functional, the increment is written as

$$\Delta\tau_{i\leftrightarrow j}^{(s,m,a)} = \left\|\mathrm{x}_j - \mathrm{x}_i\right\| * \bar{A}^{Ray}(m, \hat{\mathbf{d}}_{ij}). \tag{44}$$

The complex traveltime on each node is then relaxed in both directions according to,

$$\begin{aligned} \tau_j^{(s,m,\odot)} &\Leftarrow \min\left(\tau_j^{(s,m,\odot)}, \tau_i^{(s,m,\odot)} + \Delta\tau_{i\leftrightarrow j}^{(s,m,\odot)}\right), \\ \tau_i^{(s,m,\odot)} &\Leftarrow \min\left(\tau_i^{(s,m,\odot)}, \tau_j^{(s,m,\odot)} + \Delta\tau_{i\leftrightarrow j}^{(s,m,\odot)}\right), \end{aligned} \tag{45}$$

which constitutes a local relaxation step. Whenever an update occurs, the predecessor node index is recorded at the succeeding node, representing a segment of the raypath. After determining the minimum traveltime via local relaxation for all nodes in a current source block, the smallest traveltime is assigned as the block traveltime, given by

$$aT_{k_s}^{(s,m,\odot)} = \min_{j\in\mathcal{N}(k_s)} \tau_j^{(s,m,\odot)}. \tag{46}$$

The computation proceeds to the neighboring blocks until no unvisited blocks remain and all nodes' travel times have been updated.

Once the traveltime fields and predecessor-node indices are computed for all nodes and blocks, the discrete raypaths can be reconstructed as sequences of node IDs, following equations 41 and 42, from a source node to all receiver nodes by backtracking from the receiver nodes toward the source node. We generate a parent system consisting of parent links that interconnect the predecessor pointers during backtracking. The parent arrays $(\mathrm{par}^{(s,m,p)}, \mathrm{par}^{(s,m,a)})$ store the predecessor node IDs for each node. Referring to the notation system of equations 41 and 42, the parent pointers are equivalently written as,

$$i_{D^{(m,\odot)}}^{(s,m,\odot)} = i_r, \qquad i_{l-1}^{(s,m,\odot)} = \mathrm{par}^{(s,m,\odot)}\left(i_l^{(s,m,\odot)}\right), \tag{47}$$

for $l = i_r, i_r - 1, \ldots, i_s$. The associated real-space coordinates along the propagation and attenuation paths are expressed as

$$\mathrm{x}_q^{(s,m,\odot)} = \left(x_{i_q}^{(s,m,\odot)}, z_{i_q}^{(s,m,\odot)}\right), \qquad q = i_s, i_s + 1, \ldots, i_r. \tag{48}$$

These sequential points yield a piecewise representation of the continuous dissipative anisotropic raypaths for the first-arrival events in a 2-D model. After all paths from receiver nodes reach a source node, the computation is iteratively restarted to model the raypaths and traveltime fields for each source node location until all source nodes have been processed.

Such computation is performed independently for each wave mode and for the separate components of propagation and attenuation. Consequently, there is no overlapping (crosstalk) between wave modes—particularly shear waves—or between propagation and attenuation, which is essential for obtaining an explicit characterization. This feature is critical when the results are used as an initial model for multi-parameter tomography and/or seismic migration, where crosstalk is problematic, constituting the first significant advantage of the grid-based ray tracing scheme. Additionally, by tracing backward from receiver nodes, the method guarantees that all receivers are honored, providing a second key advantage.

## NUMERICAL EXPERIMENTS

The numerical investigation in this study is limited to OEV and C-RRT, as these are the closest options to MEV within the same shortest-path workflow. OEV provides the original g-Hamiltonian reference, while the C-RRT serves as the first conjugate eigenvectors implemented as an approximation formula to mitigate qSV cusps and triplicaiton challenges.

A comprehensive comparison of other energy-velocity formulations has been conducted at the local scale, focusing on the p-Hamiltonian and two types of c-Derivatives, demonstrating the robustness of the complex eigenvector-based construction of the energy-locity formula (Zhou et al., 2024). Another pivotal group of ray-tracing methods, including finite-difference eikonal solvers, fast marching, shooting/bending, and fast sweeping, remains admissible in the broader study, yet they employ different numerical techniques and generally do not yield the same attenuation and propagation results. Notable studies include the classical ray tracing (Cerveny, 2001) and eikonal-based solvers for traveltime modeling (Van Trier and Symes, 1990; Vidale, 1990; Podvin and Lecomte, 1991).

### Node-scale ray quantities and slowness angles

Before building the model, we focus on identifying the most robust scenarios for computing ray quantities that apply to all wave modes and run efficiently. Accordingly, we employ the published complex density-normalized elastic moduli of sandstone (Table 1) as a representative medium, which exhibits prominent triplication and cusp behavior in qSV waves due to the strong elastic anisotropies $\left(\epsilon^{(R)} \approx 0.1, \gamma^{(R)} \approx 0.08\right)$ as well as relatively strong attenuation. These characteristics significantly influence the search range and step size of the imaginary slowness-vector angle $\vartheta$ to satisfy the homogeneity condition in $\mathbf{v}$. This is one of the primary challenges in ray tracing through viscoelastic media (Zhou et al., 2024). This particular challenge originated in the earliest studies of ray tracing in such media (Vavryčuk, 2007b).

Table 1. Rock properties for the four layers of the 2D dissipative VTI model. The first layer is also used for the node-scale optimization and kernel-comparison analysis, so the table defines both the layered field model and the reference material for local ray-quantity tests.

| $\text{Layer}^{th}$ | Density-normalized Modulus $a_{pq}^{(R)}\left(\frac{km}{s}\right)^2$ | | | | | Quality Factor $Q_{pq}$ | | | | |
|---|---|---|---|---|---|---|---|---|---|---|
| | $a_{11}$ | $a_{13}$ | $a_{33}$ | $a_{44}$ | $a_{66}$ | $q_{11}$ | $q_{13}$ | $q_{33}$ | $q_{44}$ | $q_{66}$ |
| $1^{st}$ | 14.40 | 4.50 | 9.00 | 2.25 | 3.00 | 30.0 | 15.0 | 20.0 | 15.0 | 12.0 |
| $2^{nd}$ | 21.60 | 6.75 | 13.50 | 3.38 | 4.50 | 45.0 | 22.5 | 30.0 | 22.5 | 18.0 |
| $3^{rd}$ | 28.80 | 9.00 | 18.00 | 4.50 | 6.00 | 60.0 | 30.0 | 40.0 | 30.0 | 24.0 |
| $4^{th}$ | 36.00 | 11.25 | 22.50 | 5.63 | 7.50 | 75.0 | 37.5 | 50.0 | 37.5 | 30.0 |

The complex energy velocities are solved using the OEV, C-RRT, and MEV formulations based on equations 19, 22, and 27, respectively. However, since these kernels do not achieve the homogeneous energy-velocity condition defined in equation 14, iteration must be launched. This iteration searches for imaginary angles within a given search range, using small incremental steps around a given real angle, to construct the slowness direction. Then, to examine the homogeneous condition, the resulting complex energy velocity must be fed using the cost function.

In this study, the cost function used is $1-\frac{|\mathbf{v}^{(R)}.\mathbf{v}^{(I)}|}{|\mathbf{v}^{(R)}||\mathbf{v}^{(I)}|}<10^{-6}$ and since the medium is VTI, the slowness direction is determined by equation 12. Meanwhile, the search angle step is $\Delta\vartheta = 0.01^0$, with a real slowness angle range $\theta$ $0^0-90^0$. To improve computational efficiency, we have also identified four optimization functions for solving imaginary and real angle pairs, namely *fminunc*, *fsolve*, *fminsearch*, and *fminbnd* in MATLAB (Table 2). These functions were embedded into each kernel (OEV, C-RRT, and MEV) and compared with the direct iterative formulation. The first three functions solve the required $\vartheta$ using unconstrained conditions, while the last one is a bounded-angle minimization function. We used a wide range of $\vartheta$ $(-19^0<\vartheta<19^0)$ to test the robustness for all body waves. Without implementing the

optimization functions, the iterative approach can take more than 12 minutes to calculate all qualified ray quantities for all wave modes at a single node (Table 2). However, by utilizing these functions, the runtime is reduced by several orders of magnitude compared to the iterative scheme. In general, such improvements scale favorably, yielding greater exponential acceleration when multiplied by the number of nodes.

The optimization function study is not included merely as an implementation detail. In the present method, the local kernel is iteratively computed across angular samples, wave modes, nodes, and blocks, and for both propagation and attenuation. As a result, the runtime of solving the Hamiltonian subproblem at nodes escalates multiplicatively with the number of nodes, blocks, wave modes, and angular samples, increasing the overall elapsed time on both nodes and at the field scale for ray tracing. The selection of robust optimization methods is therefore part of the fundamental framework, not merely an optional practice.

Among the four functions, *fminbnd* consistently performs the fastest, with a runtime almost half that of the three other functions, followed by *fminsearch*, with only minor runtime differences (Table 2). Therefore, we select *fminbnd* to optimize the energy-velocity formulation and further analyze the reliability of MEV relative to C-RRT and OEV.

Substantial acceleration is crucial not only as a computational advance in the formulation of energy velocities but also in facilitating the exponential decrease in runtime during local and global ray tracing updates in our methodology. This decrease is due to the kernel function being called multiple times per block, a characteristic of this grid-based approach. Furthermore, efficient and accurate ray tracing is essential, as it serves as a wave solver during tomography, where it is iteratively called to update the traveltime and raypath for accurate velocity modeling.

Table 1. Elapsed time (s) for iterative and optimized kernels used to compute complex energy velocities with the three formulations. The fastest values are highlighted in bold, showing that the bounded one-dimensional search provides the best compromise between robustness and runtime.

| | Iterative(s) | *fminbnd* (s) | *fminunc* (s) | *fminsearch* (s) | *fsolve* (s) |
|---|---|---|---|---|---|
| OEV | 241.237 | **0.487** | 1.049 | 0.897 | 1.123 |
| C-RRT | 256.957 | **0.569** | 1.031 | 0.914 | 1.124 |
| MEV | 736.410 | **1.946** | 3.766 | 3.293 | 4.032 |

We further identify the most robust kernel to build the complex energy velocity and the subsequent ray quantities. Cusps in the qSV wavefront for this parameter set have already been documented, so we do not emphasize 2D wavefront plots where the cusp is obvious. Instead, we analyze scalar ray quantities and ray angles as functions of the real slowness (take-off) angle $\theta$ to maximize sensitivity to subtle differences between the kernels. In addition, the plotting configuration in Fig. 2 yields explicit angular variations in the ray quantities, highlighting the possible propagation behavior of each wave within the model.

In the kernel comparison analysis, we use the OEV as the original real-valued ray tracing kernel and the benchmark model for qP and qSH waves. The OEV reliably reproduces its ray quantities and imaginary angles, as demonstrated by the undistorted, smooth curves. However, it yields distorted qSV ray-quantity manner when the cusps are present (Fig. 2), as indicated by distorted values as the real slowness angles $\theta$ change within the ranges $20^0 - 30^0$ and $45^0 - 50^0$. At these angles, the imaginary-angles $\vartheta$ abruptly amplified ($\vartheta > 10^0$ & $\vartheta < -10^0$). This behavior is consistent with previous studies that identify the challenge of the OEV in dealing with cusps on qSV (Li et al., 2020; Wu et al., 2021b; Zhou et al., 2024).

In contrast, C-RRT, whether implemented iteratively or with optimization, exhibits substantial deviations in qP ray quantities and in the search for $\vartheta$. This behavior is shown by distinct abrupt variations of $45^0 - 50^0$ slowness vector angles $\theta$, where at these points the angles $\vartheta$ abruptly changed ($\vartheta > 5^0$ & $\vartheta < -10^0$). In addition, this kernel also fails to reliably model qSV wave ray quantities and slowness angles $\theta$, with $\theta$ distortions occurring around $50^0 - 55^0$ and $65^0 - 70^0$, where at these points the angles $\vartheta$ are significantly perturbed, similar to those observed for OEV ($\vartheta > 10^0$ & $\vartheta < -10^0$). These results imply that, within our parameter regime, the C-RRT solution is much less robust than the MEV solution. The results also indicate the drawbacks associated with neglecting the asymmetric condition of the Hamiltonian matrices. The MEV kernel appears to be the only approach that consistently models all body-wave ray quantities and angles. The qP and qSH ray quantities from MEV agree exactly with those from OEV, in accordance with theoretical expectation, while MEV outperforms the other kernels in the qSV case. In the following section, further comparisons between the MEV, C-RRT, and OEV are conducted at the global field level, focusing on raypaths and travel times. The results also provide physical insight into the raypath behavior for each wave mode. The $V_{qP}^{ray}$ and $V_{qSH}^{ray}$ increase significantly at $\theta$ angles above $40^0$, and reach a plateau from $70^0$ to $90^0$. These results indicate that the qSH and qP propagation raypaths prefer near-vertical directions as the shortest paths. Meanwhile, the $V_{qSH}^{ray}$ values peak within the angle range $20^0 - 60^0$, indicating that qSV propagation raypaths favor more diagonal trajectories. In contrast, the attenuation quantities $A^{ray}$, for all wave modes, vary only subtly, on the order of hundredths. Consequently, even slight variations in controlling factors, such as propagation distance or rock properties, can trigger significant changes in attenuation trajectories.

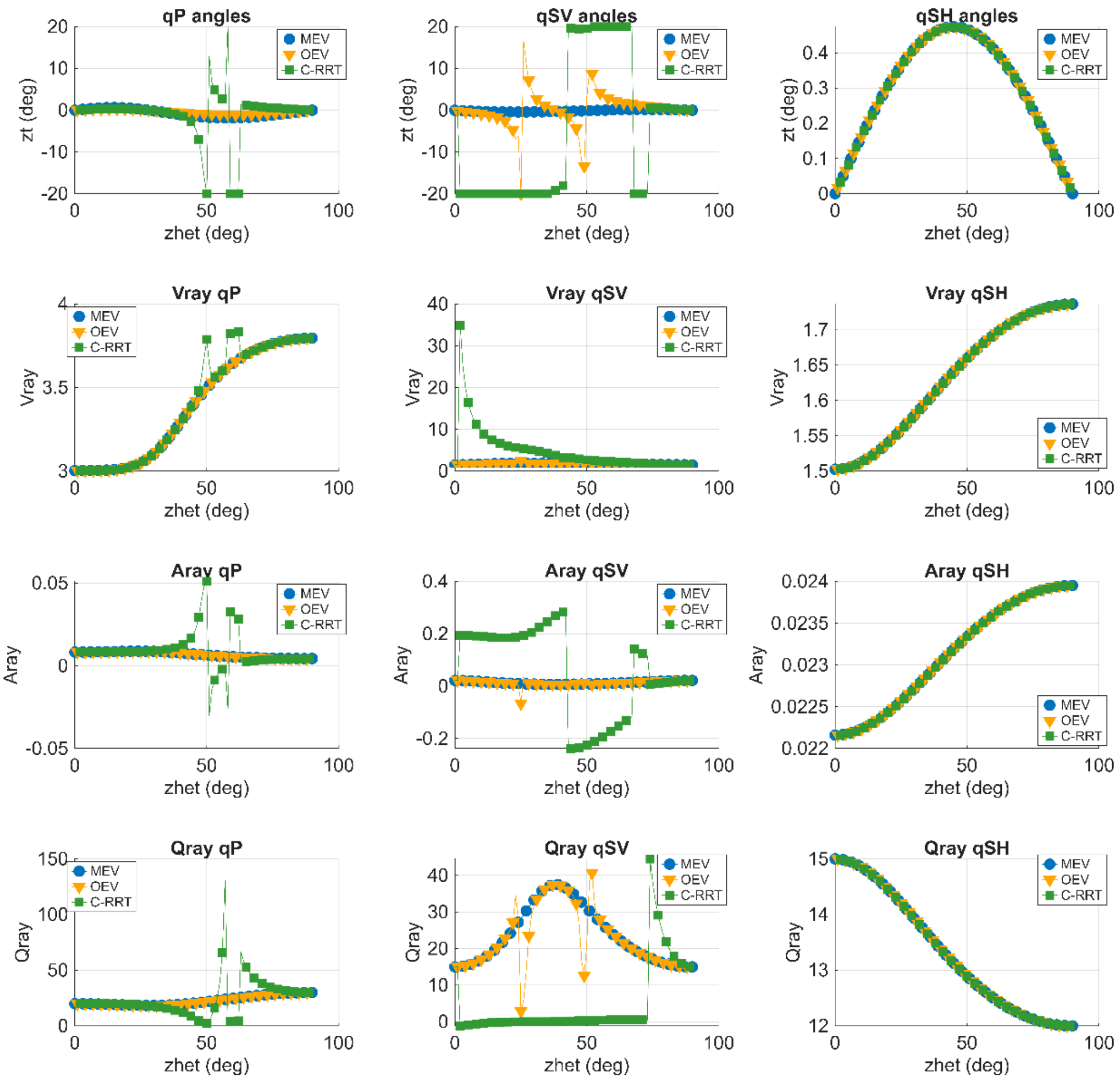

Fig. 2. Local comparison of slowness-vector angles and ray quantities computed with OEV (yellow triangles), C-RRT (green squares), and MEV (blue circles). OEV and MEV coincide for qP and qSH, whereas OEV breaks down for qSV cusps, and C-RRT shows abrupt deviations even for qP; the figure therefore identifies MEV as the most stable kernel among the tested formulas for all body waves in this parameter set.

The scalar RMS misfits in Fig. 3 are calculated for each kernel and serve as quantitative measures of discrepancies between OEV and MEV, and between OEV and C-RRT, for slowness angles and ray quantities. For qP wave parameters, OEV–MEV misfits (blue curves) are essentially at near numerical precision, demonstrating that our MEV implementation is quantitatively equivalent to OEV across all angular ranges of $\theta$. On the contrary, the C-RRT–OEV are consistently 1–4 orders of magnitude larger than those between OEV and MEV, especially $A^{Ray}$ and $Q^{Ray}$, with the highest discrepancies occurring at diagonal angles, as reflected by the distorted surfaces (Fig. 2). For the qSV wave, all kernel discrepancies vary heavily, as expected, around the cusps or triplications, yet preserve the same pattern as qP. In this case, MEV remains closely coherent with OEV, as indicated by the moderate RMS errors, while C-RRT continuous to produce substantial errors, particularly in $A^{Ray}$ and $Q^{Ray}$. The qSH mode across all kernels is nearly identical and hence trivial, as the RMS are negligible.

These RMS results provide two key insights. First, the qP and qSH ray quantities computed from OEV can serve as a benchmark for assessing improvements in advancing Hamiltonian formulation to overcome cusps or triplications in qSV. Second, they also provide additional evidence for a systematic weakness in the C-RRT formulation, as the model can be misleading even in the absence of cusps, i.e., for qP results. This deficiency is further demonstrated by its large RMS values in the attenuation parameters, which can abruptly amplify moderate ray-velocity errors, especially at intermediate angles. This behavior is consistent with its wavefront models (Fig. 2), in which the homogeneous slowness vector condition cannot be fully enforced, leading to convergence toward spurious stationary points. These findings provide early validation of MEV's efficacy, while the remainder of this study focuses on the performance of these kernels in producing physically favorable first-arrival raypaths and traveltimes.

## Model Construction

Based on these ray-quantity results, we define four purely flat layers in a 2D VTI dissipative earth model (Fig. 4). Each layer contains the same sandstone properties, scaled by a multiplicative factor (Table 1). The total model dimensions are 2880 m in the lateral direction and 640 m in the vertical direction, with a uniform layer thickness of 160 m. To provide the model with an even distribution of directional angles, all blocks are rectangular, ensuring equal ray coverage. These blocks are constructed with dimensions of 160m × 160m and 17 nodes on both vertical and horizontal edges. The former choice emphasizes the method's efficacy with one node per layer thickness, while the latter provides sufficient take-off angle resolution for each block ($\leq 3^o$). This resolution is significantly higher than that used in previous studies, which employed only 4–9 nodes per edge (Zhou and Greenhalgh, 2005, 2006; Bai et al., 2007; Li et al., 2020; Wu et al., 2021a). A surface-acquisition geometry is adopted, with both sources and receivers located at the surface. The receiver spacing follows the 160 m block dimension, avoiding excessively dense raypaths, improving interpretability, and reducing computational cost. Multiple source locations were deployed and parallelized across multiple GPUs. Only one representative surface source location is shown here to highlight first-arrival events, particularly refraction events.

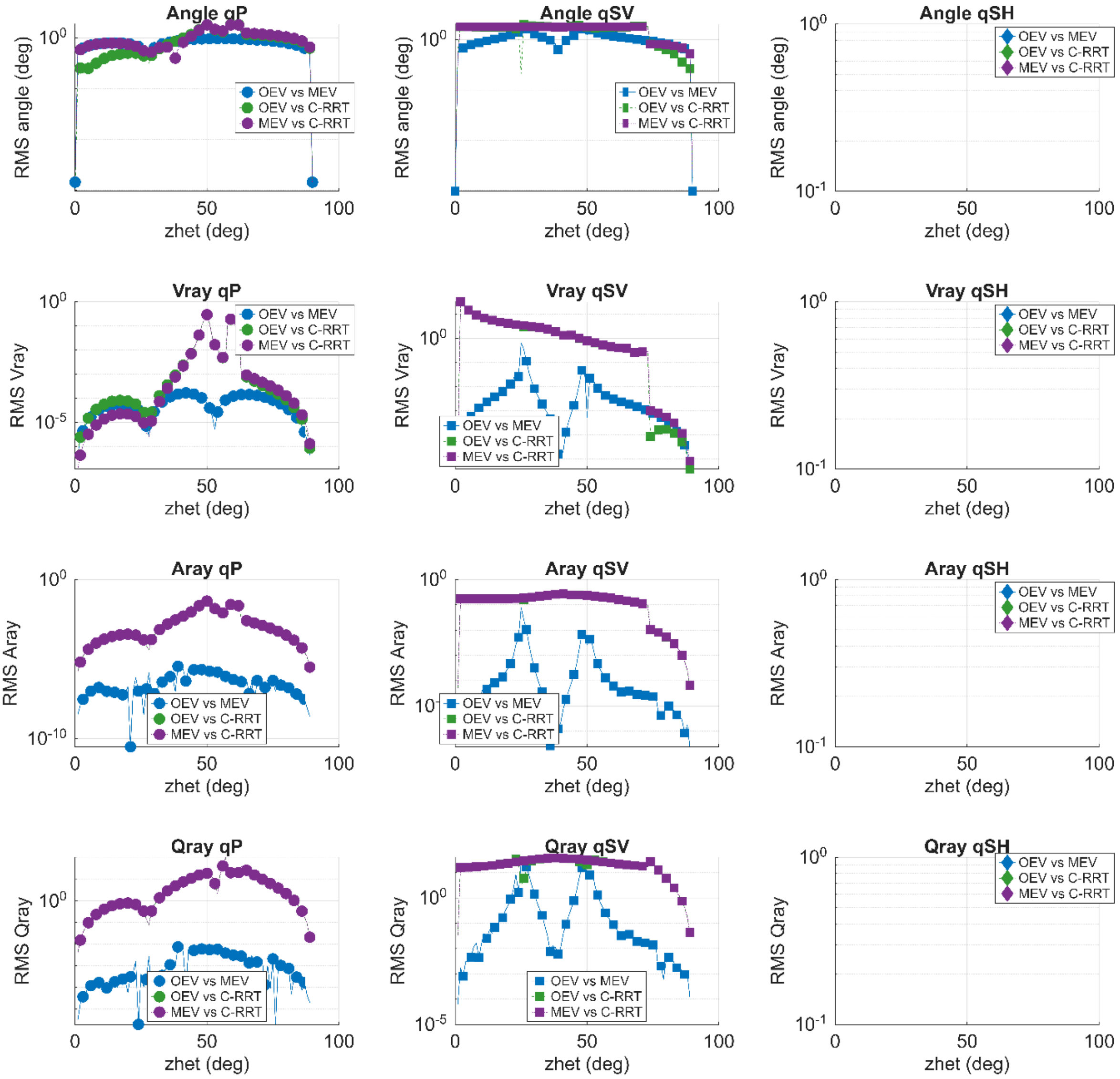

Fig. 3. RMS misfits for slowness-vector angles and ray quantities between OEV and MEV (blue dots), and between OEV and C-RRT (green dots), calculated from the trends seen in Fig. 2. OEV and MEV are nearly identical for qP and qSH and remain coherent for qSV, whereas C-RRT produces much larger errors, especially in ray attenuation and quality factor.

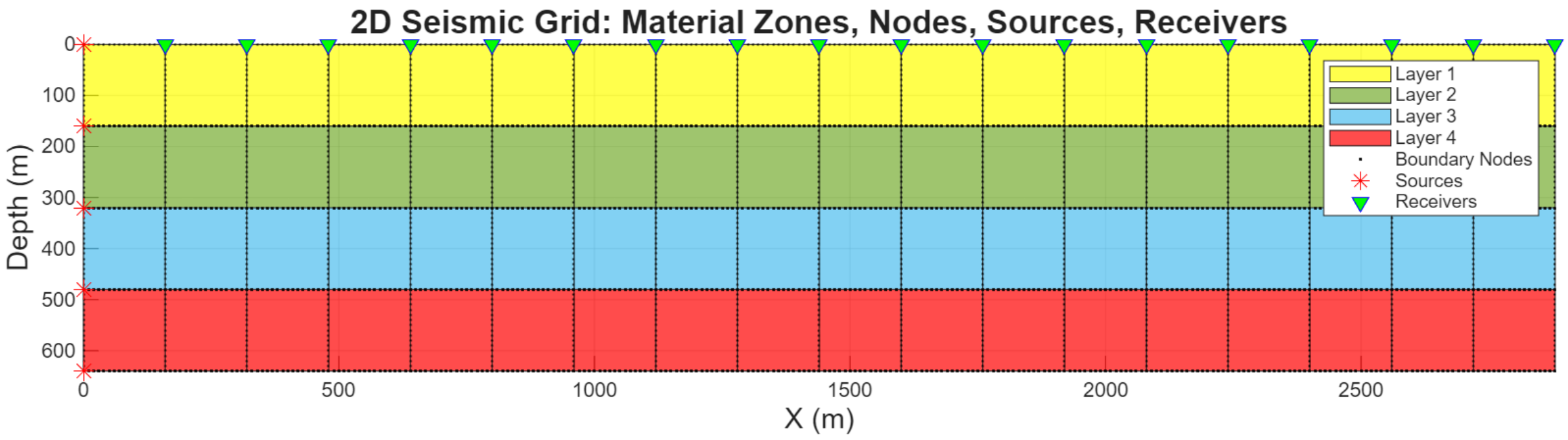

Fig. 4. Discretized-layered 2D dissipative VTI model used for field-scale ray-tracing tests, showing the discretization, receiver positions (green triangles), and source locations (red stars) along the surface and borehole. The model is constructed to execute first-arrival propagation, refraction, and attenuation in a controlled, heterogeneous setting.

### Traveltime and attenuation fields

Here, we further discuss how the distinctions between OEV, C-RRT, and MEV at the local (nodal) kernel stage translate into the field-scale model. The surface sources are located in the left part of the model, as shown by the initial traveltime contour. Based on the previous kernel-level comparisons and RMS analyses, the OEV traveltime results serve as the reference model for qP and qSH, while a standalone comparison is used for qSV traveltimes. Hence, for clarity, in this section, we directly perform the quantitative analysis by presenting difference maps for qP and qSH waves superimposed on the OEV traveltime contour.

In the qP modeling, there are negligible quantitative differences, reflected in the extremely weak amplitudes in the difference maps (Figs 5a and 5b). Thus, the propagation traveltime values, contours, spacing, and interactions with the layered media are identical for both OEV and MEV. This finding confirms that the previously observed small misfits between the two kernels in the $V^{Ray}$ and $\theta$ (Fig. 3) do not result in any discernible field-scale bias. In contrast, the difference map between the OEV and C-RRT (Fig. 5c) reveals systematic discrepancies in both the near- and far-offset regions, where velocity and attenuation effects are most potent.

These areas coincide with the angular ranges where the C-RRT misfits peak, indicating that kernel-level errors at certain $\theta$ are directly manifested as spatial perturbations in the traveltime field.

The qP attenuation expressed in traveltime contours is even more sensitive to kernel selection, consistent with the RMS results for $A_{qP}^{Ray}$ and $Q_{qP}^{Ray}$. The MEV-OEV difference map shows a coherent pair, with a pattern similar to that in the propagation field: the attenuation fields increase with depth and along longer trajectories. Their differences remain negligible, weak, and spatially smooth (Figs 6a and b). However, the C-RRT kernel produces misleading attenuation results, including negative attenuation values over the field-scale domain and amplitude amplification rather than attenuation. Such behavior is non-physical under the assumed viscoelastic model and is likely due to numerical inconsistencies arising from the approximate Hamiltonian formulation. As all layers in the model are explicitly parameterized with positive Q values without any amplitude gain mechanism, negative $\boldsymbol{\tau}^{(I)}$ results directly imply $e^{-\omega\boldsymbol{\tau}^{(I)}} > 1$, meaning that the amplitude grows along the path. Within the given model, this is clearly non-physical and is therefore considered a numerical artifact of the kernel or optimization function rather than a bona fide propagation effect. These artifacts suggest that the C-RRT formulation does not consistently enforce the homogeneous complex energy-velocity condition required for physically admissible attenuation modeling (Fig. 6c). As a result, the C-RRT–OEV difference maps exhibit pronounced patches of under- and over-attenuation, frequently aligned with regions of complex wavefront geometry, where anisotropy and viscoelasticity exert the most decisive influence (Fig. 6d). Such behavior is likely attributable, at least in part, to neglecting the asymmetry of the Hamiltonian matrices in the C-RRT formulation. To our knowledge, these specific attenuation artifacts have not been reported in previous applications of C-RRT.

The propagation traveltime fields of qSV waves emphasize the distinctive field-scale behavior of the three kernels. The OEV (g-Hamiltonian) and MEV (g*-Hamiltonian) produce similar contour patterns (Figs 7.a and 7.b), indicating that the local misfits do not translate into any field-level bias. Accordingly, their difference map remains weak and smooth, while still reflecting the overall contour geometry due to differences in traveltime magnitude, with MEV traveltimes differing from OEV by only tenths of a second. In contrast, the C-RRT solution produce signifficantly lower traveltime values than OEV and MEV, in addition to having inconsistent contour spacing in the middle offsets and depths (Fig. 7c). As the corresponding kernel behavior does not support these low traveltime values and is later accompanied by an attenuation anomaly, we consider them a numerical irregularity rather than physically genuine faster propagation. Consequently, the C-RRT–OEV difference map is not only predominantly negative, but also exhibits larger magnitudes, generating coherent traveltime-misfit patches that intensify in regions where the qSV kernel RMS are largest. These differences suggest that C-RRT fails to maintain homogeneous qSV energy velocities at the field scale, particularly where the wavefront is most asymmetric, thereby introducing regional, yet systematic, biases.

For the qSV attenuation fields, the contours naturally reflect the complex kinematics through irregular geometries (Fig. 8). Nevertheless, they exhibit a hierarchy similar to that observed for qP attenuation, with OEV and MEV producing comparable contour patterns. However, C-RRT again yields negative attenuation values, accompanied by patchy anomalies and irregular transitions, with no corresponding features in the OEV or MEV results. Meanwhile, the qSH propagation traveltime fields and attenuation in traveltime units serve as an additional control, as all kernels generate nearly identical contours, consistent with near-zero RMS and difference maps (Figs 9 and 10). Their contour characteristics are also similar to qP, revealing systematic variations in the near- and far-offset regions where both velocity and attenuation effects are most potent.

Overall, these field-scale results imply that the minor kernel-scale discrepancies between OEV and MEV (Fig. 3) do not propagate into significantly large-scale differences, whereas the larger C-RRT misfits, particularly in the regime of $A^{Ray}$ and $Q^{Ray}$, do. Regions where OEV — MEV — C-RRT differs most strongly are closely correlated with the angular ranges where C-RRT application does not conserve a stable homogeneous qSV energy-velocity at the local scale. This observation supports the hypothesis that neglecting the Hamiltonian asymmetry leads to physically misleading attenuation behavior in the traveltime field. At this stage, both MEV and OEV, when optimized using fminbnd, produce physically plausible complex traveltime fields; the study, therefore, involves a comparative analysis of raypaths.

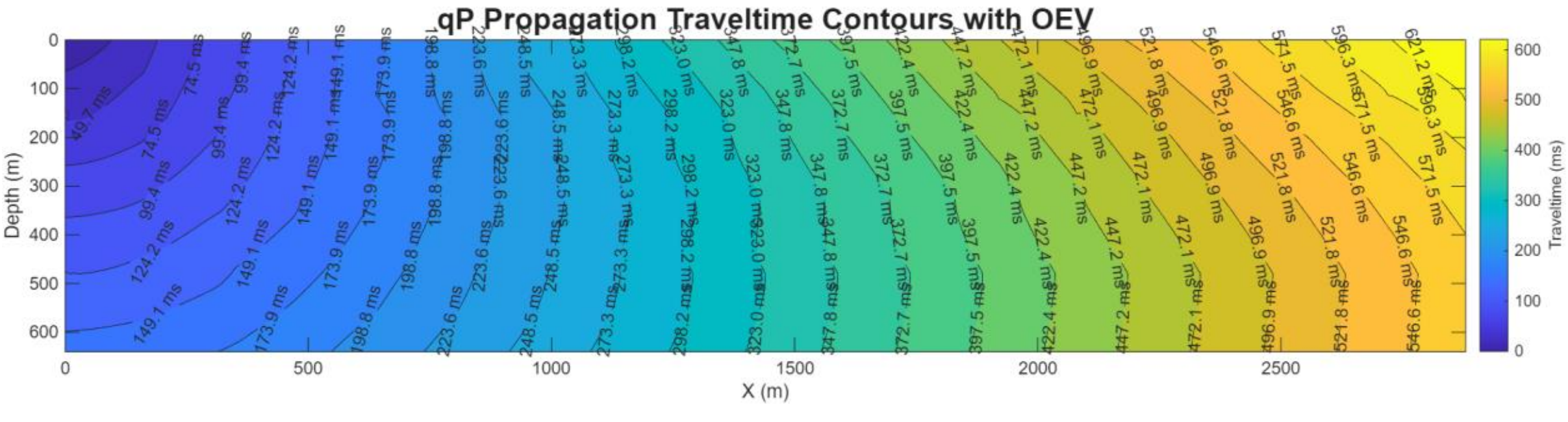

*a.*

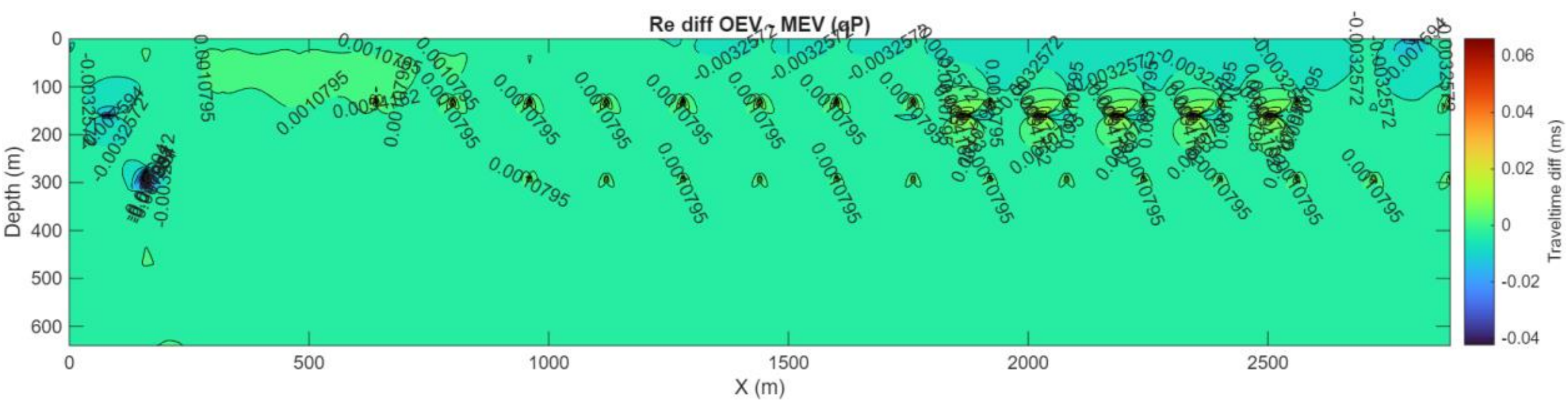

*b.*

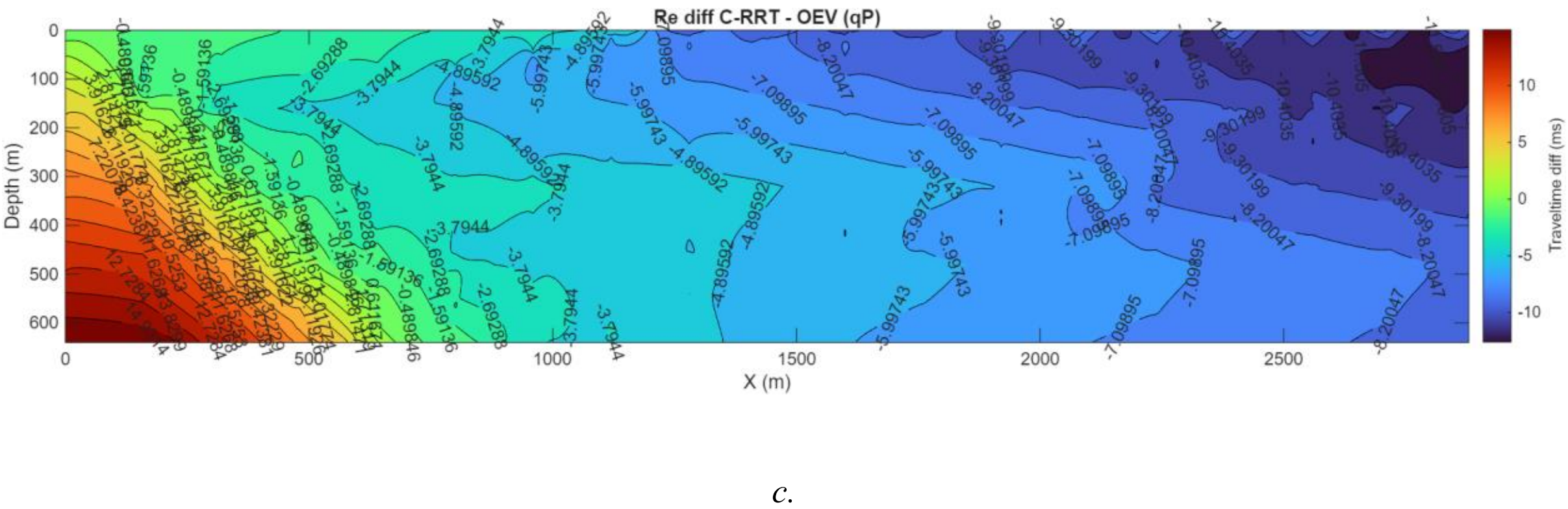

*c.*

Fig. 5. qP propagation-traveltime contours for (a) OEV, with traveltime difference maps relative to (b) MEV and (c) C-RRT. OEV and MEV results are essentially indistinguishable at the global scale, whereas C-RRT produces coherent traveltime biases at near and far offsets, as well as deeper depth.

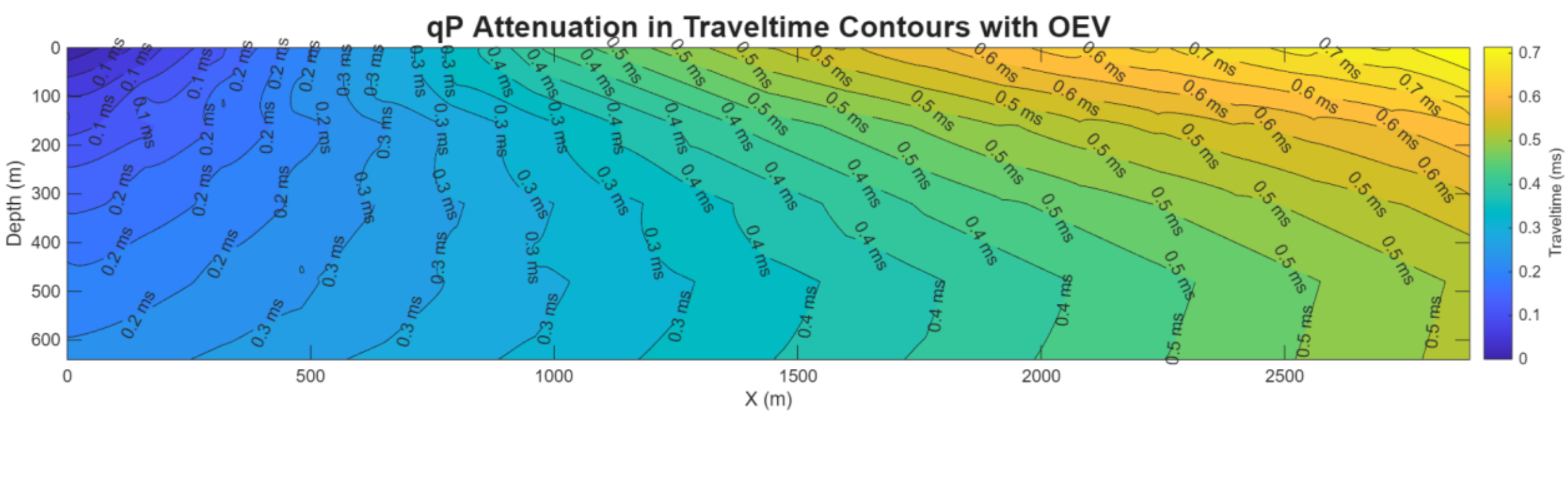

*a.*

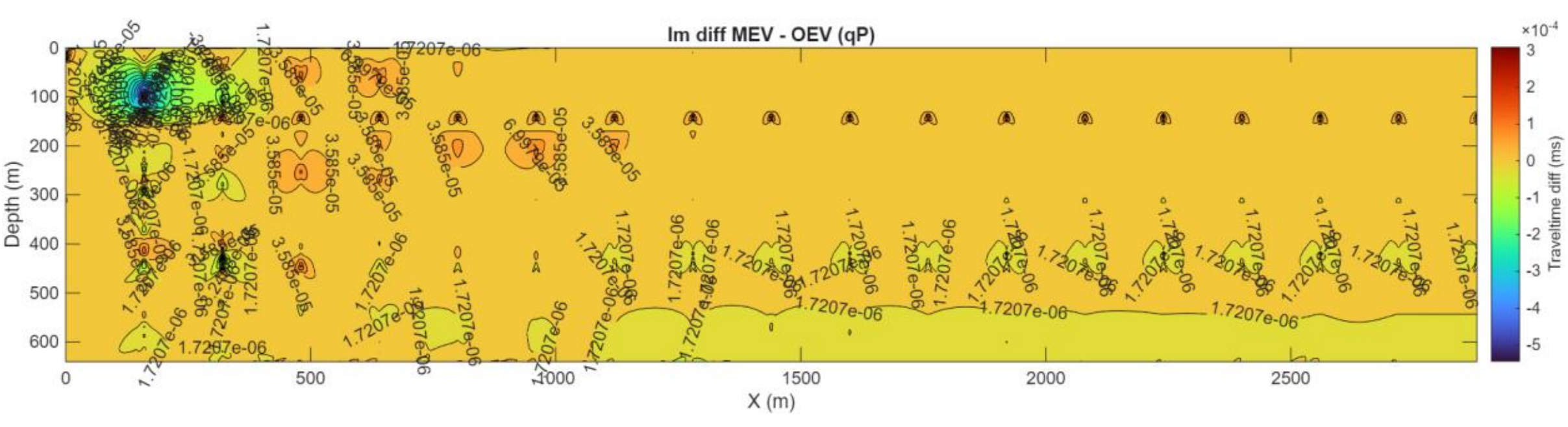

*b.*

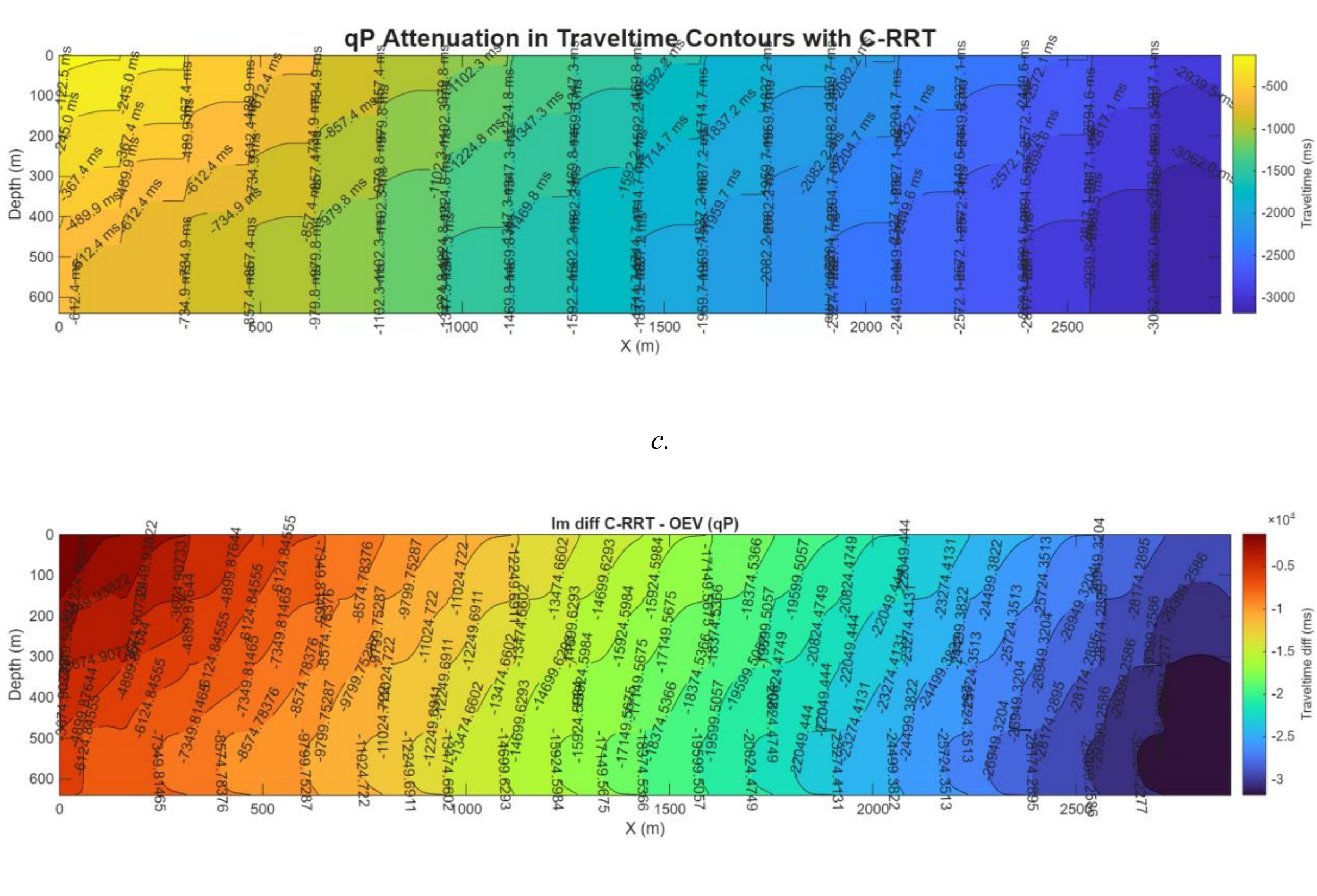

*c.*

*d.*

Fig. 6. qP attenuation expressed in traveltime units for (a) OEV, with the OEV-MEV traveltime difference in (b), the C-RRT field in (c), and the OEV-C-RRT traveltime difference in (d). OEV and MEV produce indistinguishable attenuation fields, while C-RRT produces nonphysical negative attenuation values that indicate amplification artifacts.

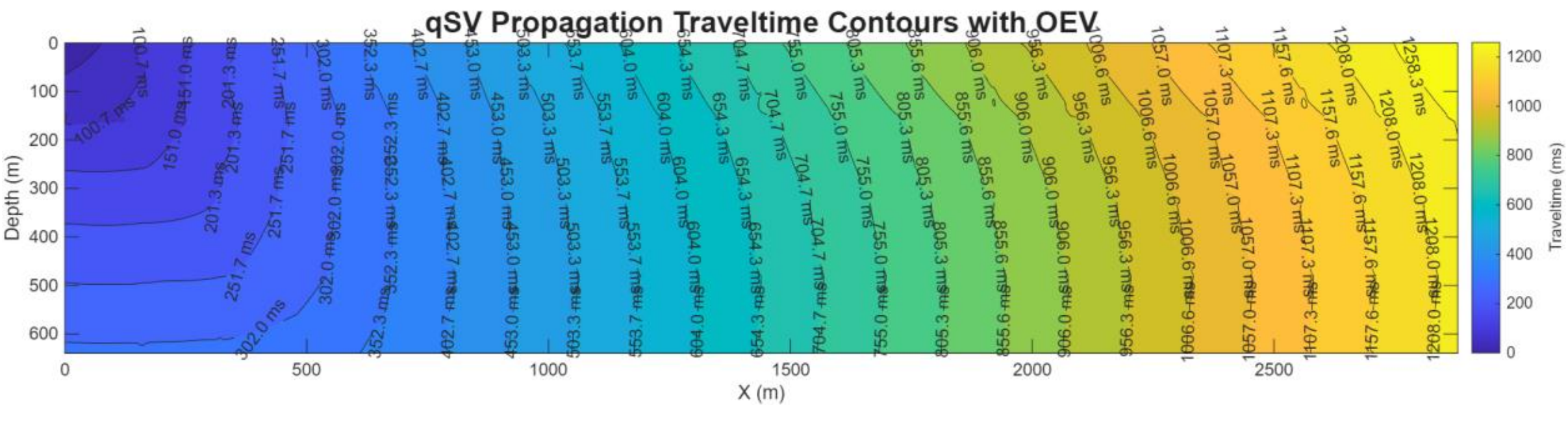

*a.*

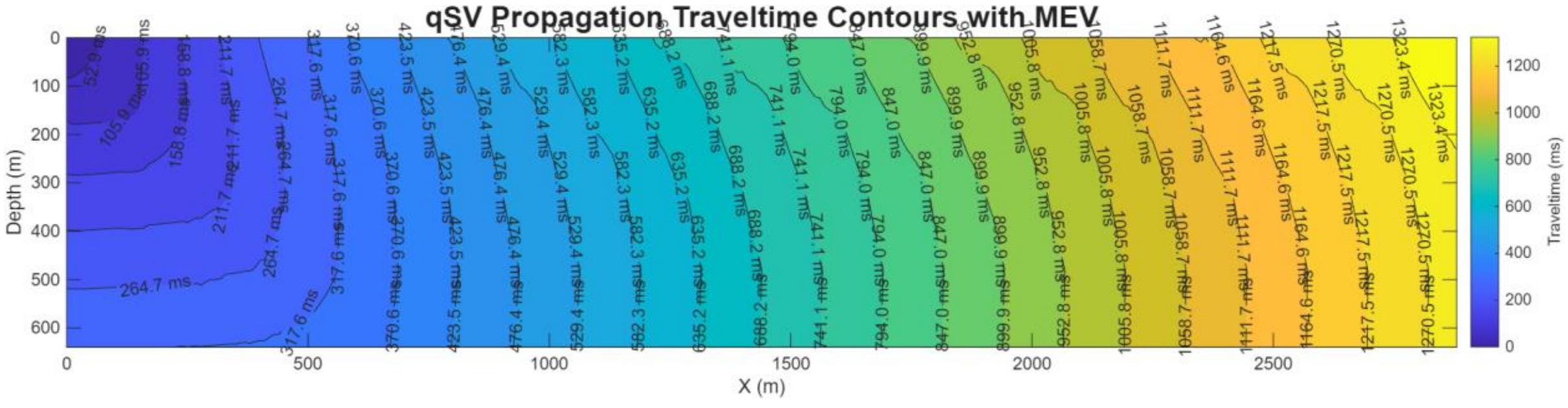
qSV Propagation Traveltime Contours with MEV
Depth (m)
X (m)
Traveltime (ms)

*b.*

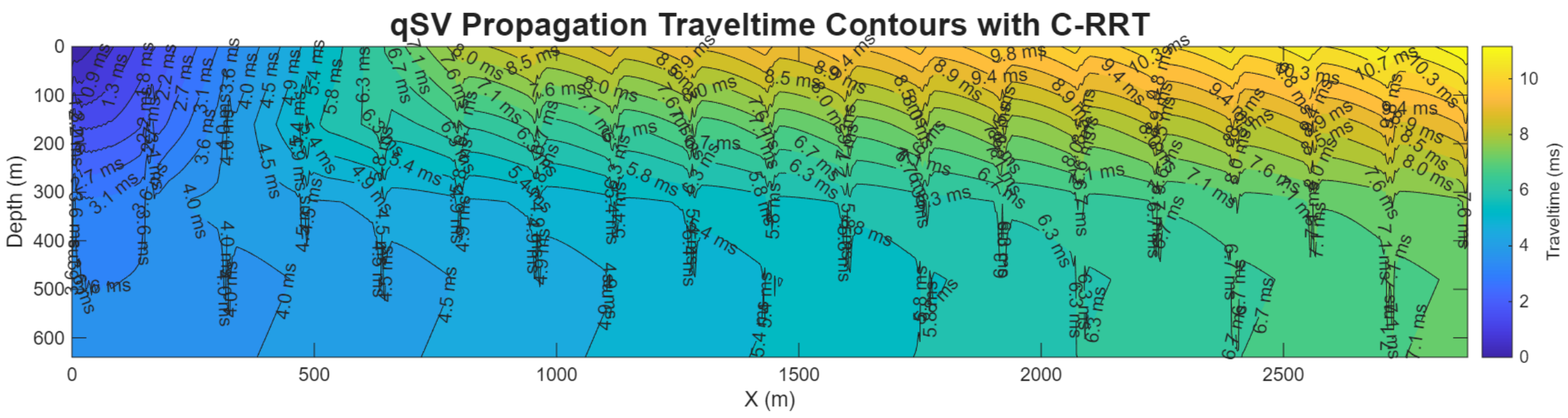
qSV Propagation Traveltime Contours with C-RRT
Depth (m)
X (m)
Traveltime (ms)

*c.*

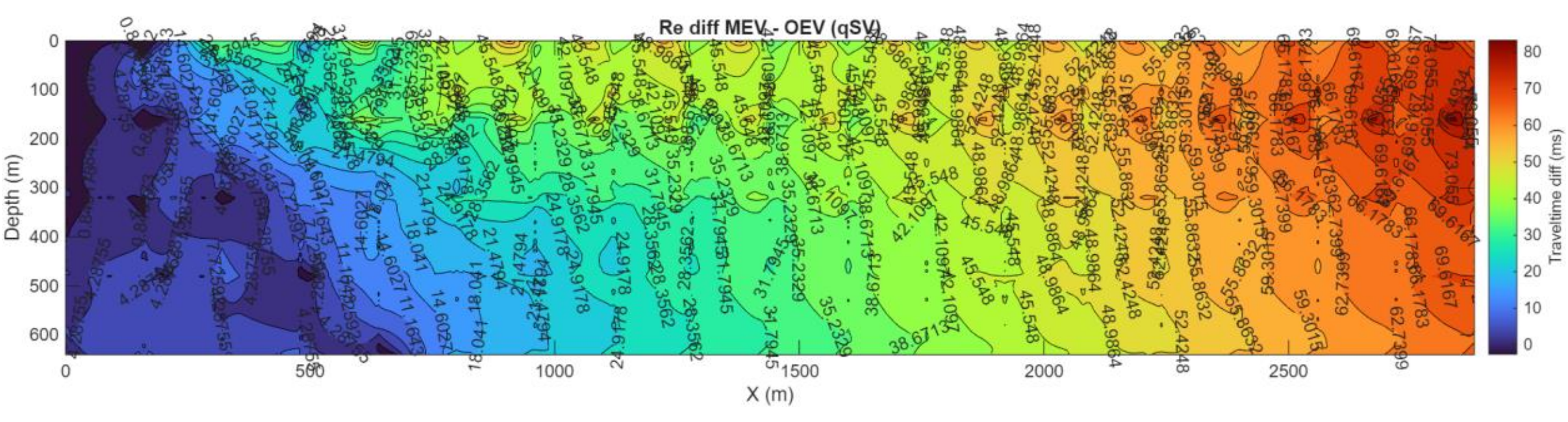
Re diff MEV - OEV (qSV)
Depth (m)
X (m)
Traveltime diff (ms)

*d.*

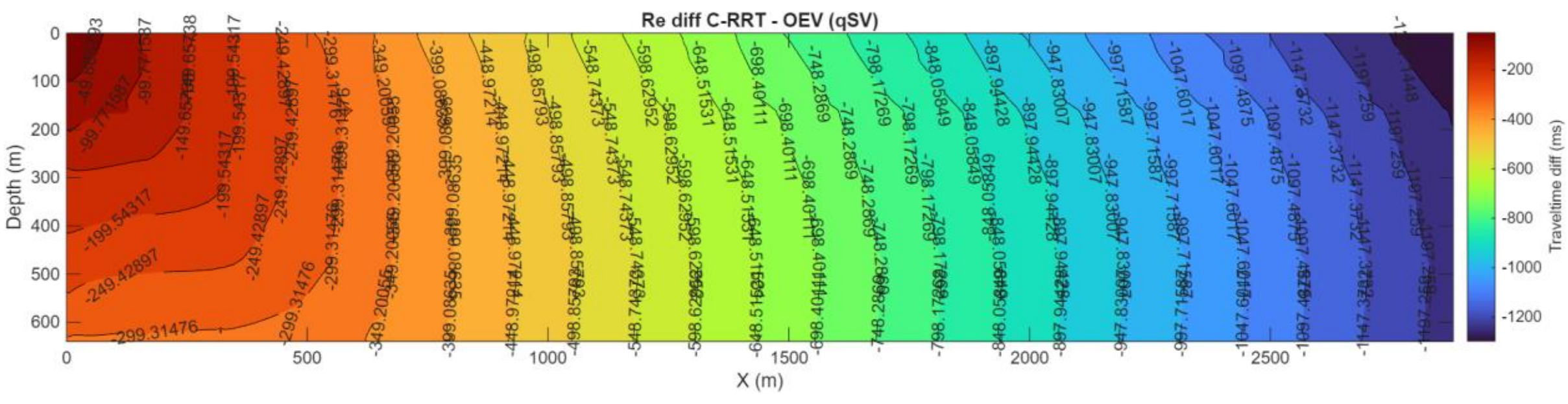
Re diff C-RRT - OEV (qSV)
Depth (m)
X (m)
Traveltime diff (ms)

*e.*

Fig. 7. qSV propagation traveltime contours of (a) OEV, (b) MEV, and (c) C-RRT solutions, with traveltime difference maps of (d) OEV-MEV and (e) OEV-C-RRT. Both MEV and OEV exhibit similar behavior, whereas the C-RRT solution yields noticeably lower values and larger coherent misfits.

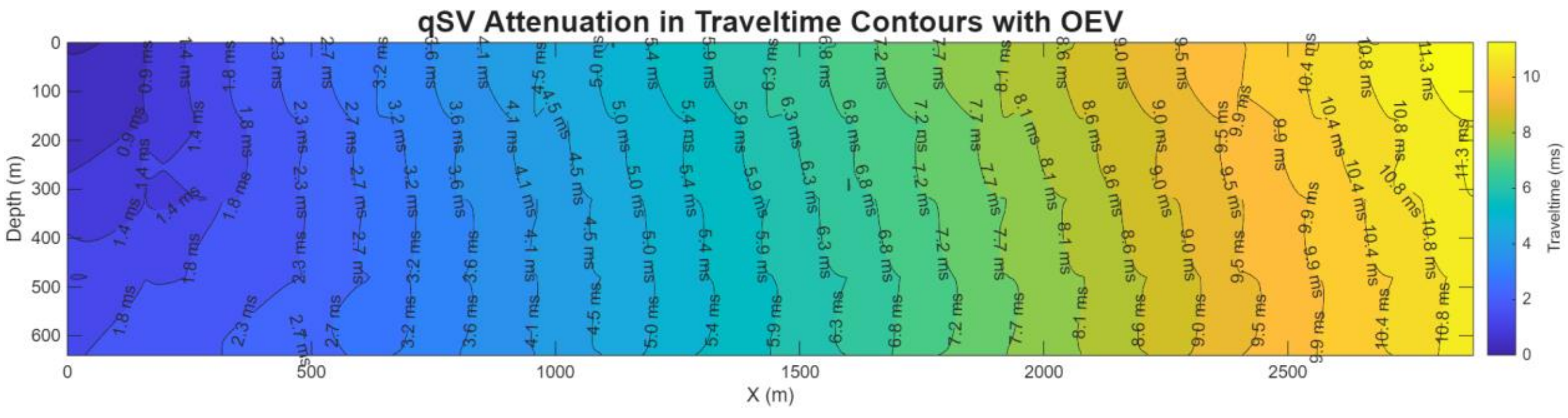

*a.*

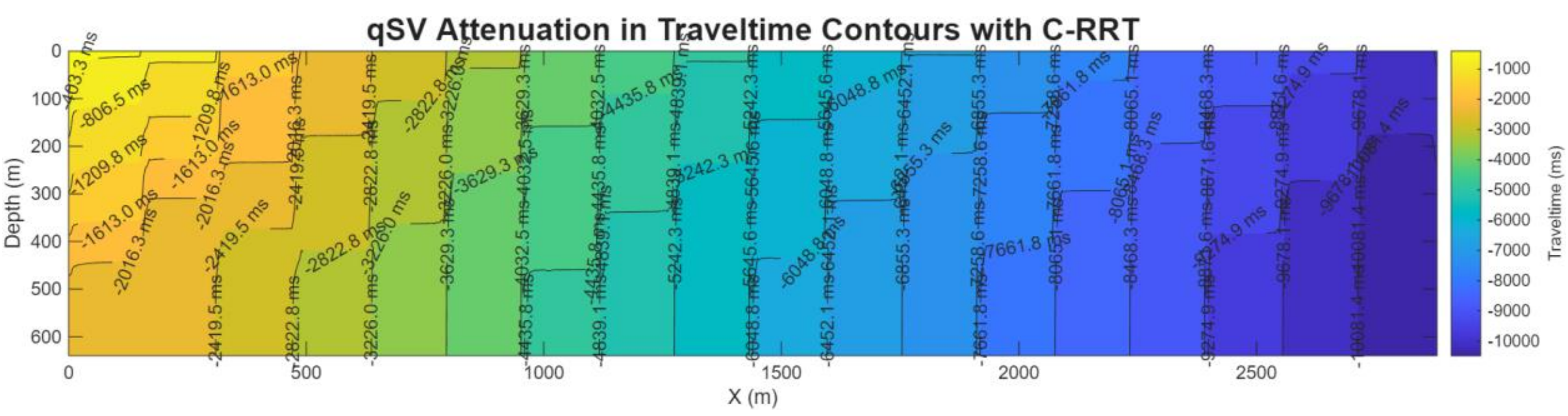

*b.*

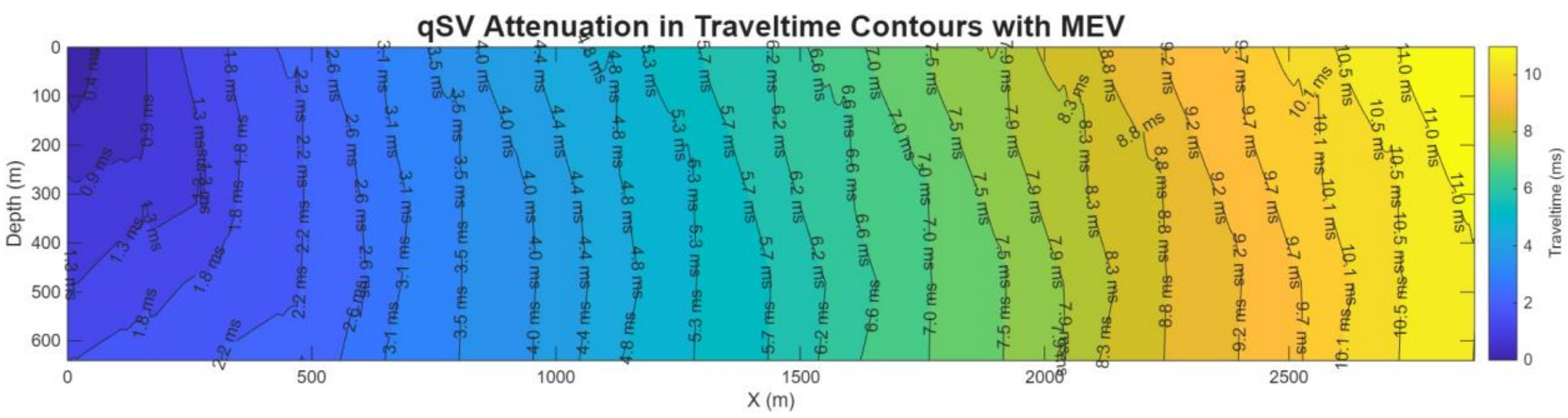

*c.*

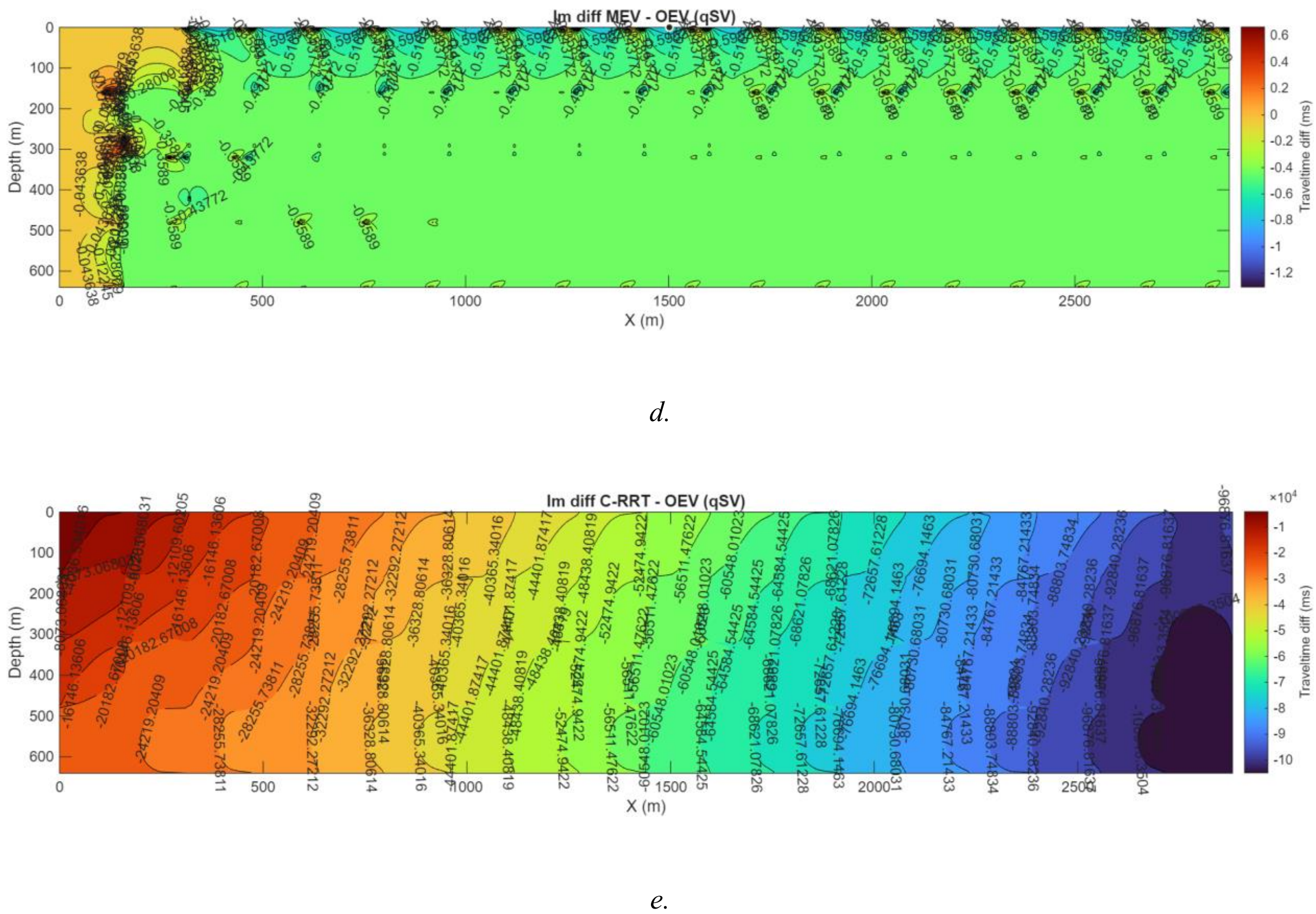

*d.*

*e.*

Fig. 8. qSV attenuation in traveltime units for (a) OEV, (b) C-RRT, and (c) MEV, with traveltime difference maps (d) OEV-MEV and (e) OEV-C-RRT. OEV and MEV remain coherently consistent, while C-RRT yields localized anomalies and negative attenuation, indicating numerical instability in qSV modeling.

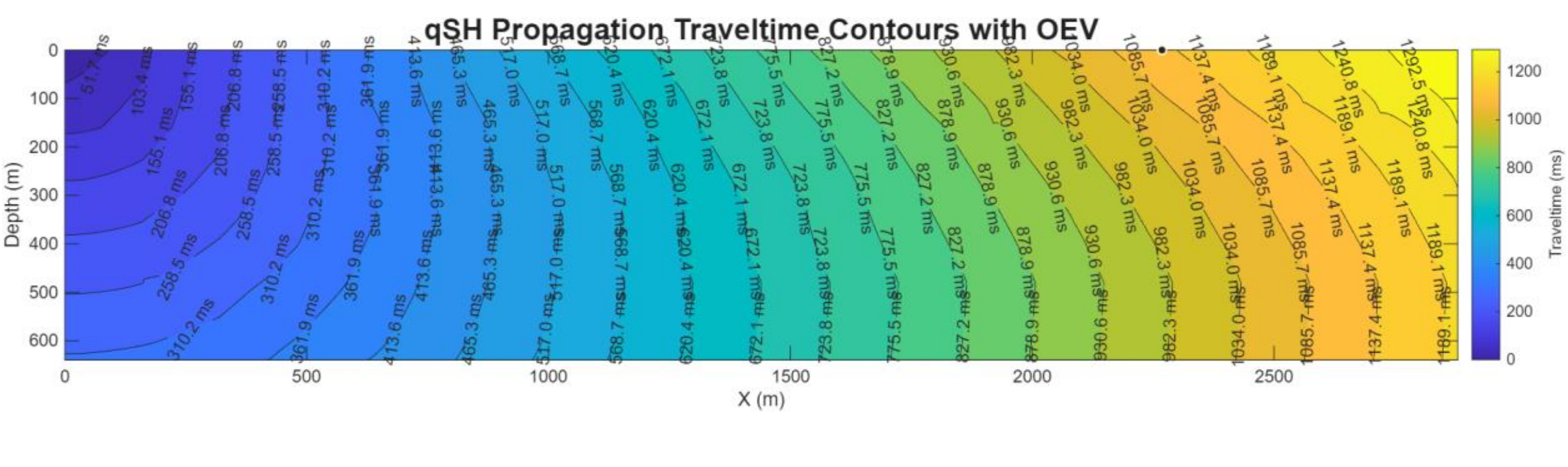

*a.*

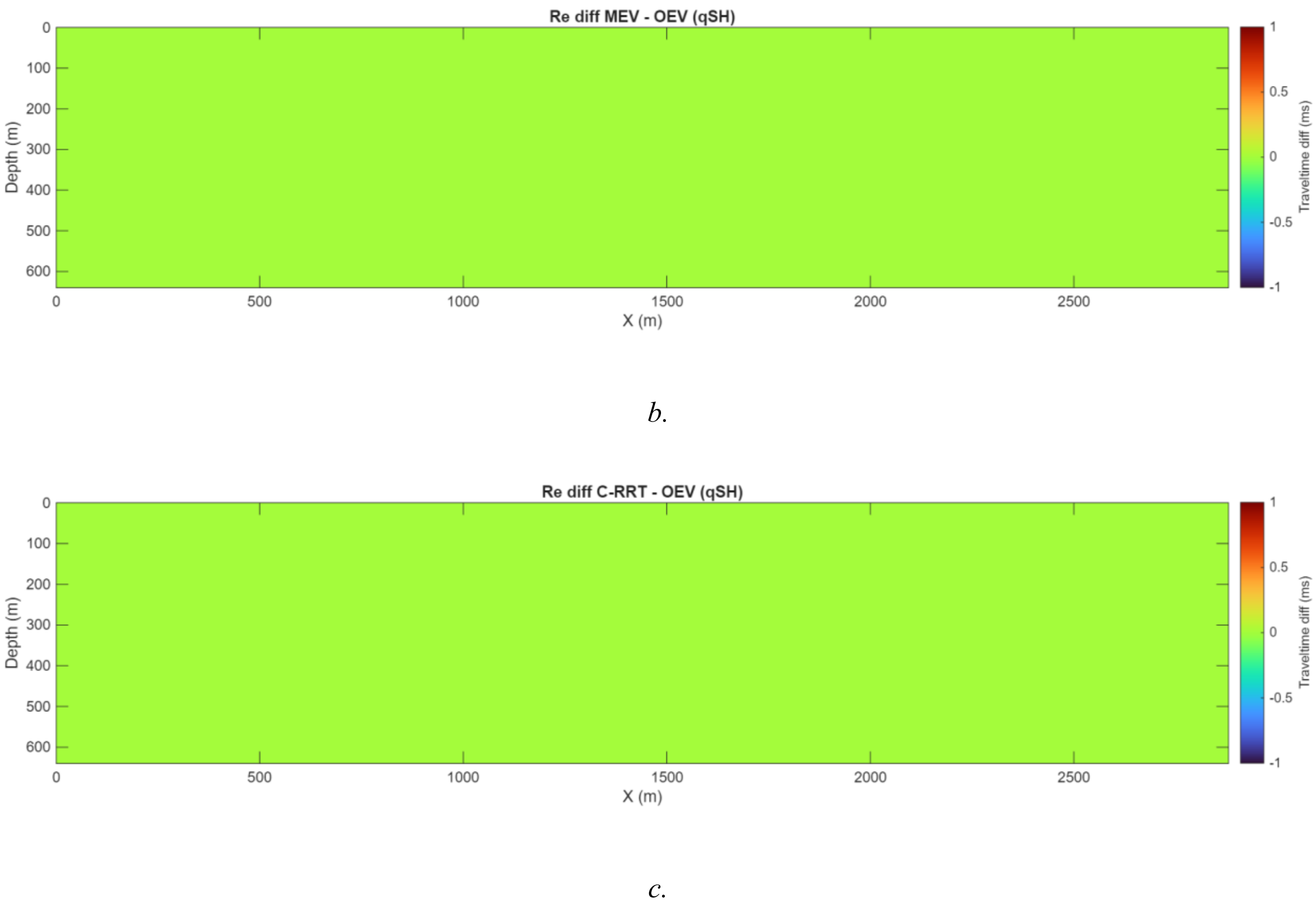

*b.*

*c.*

Fig. 9. qSH propagation-traveltime contours for (a) OEV, with traveltime difference maps (b) OEV-MEV and (c) OEV-C-RRT. All three kernels are completely similar for qSH, confirming that the differences observed for qP and qSV are not generic numerical artifacts.

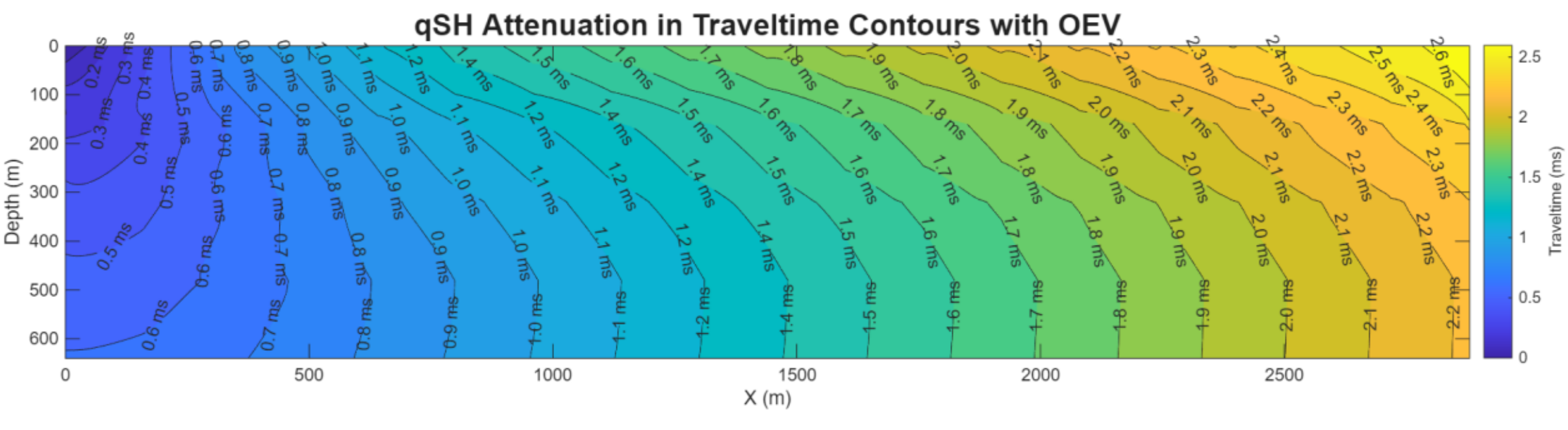

*a.*

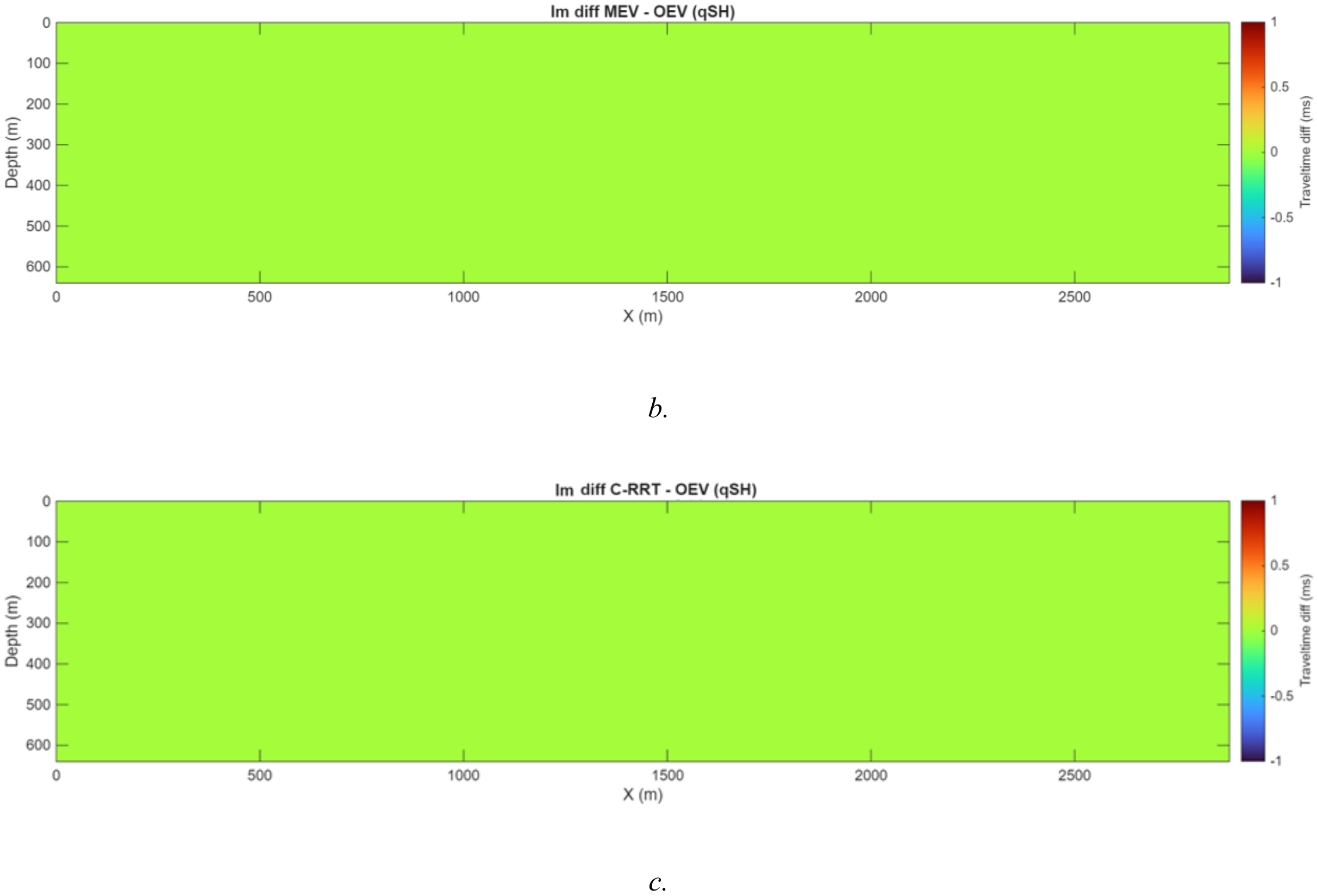

*b.*

*c.*

Fig. 10. qSH attenuation in traveltime units for (a) OEV, with traveltime difference maps relative to (b) OEV-MEV and (c) OEV-C-RRT. All three kernels are completely similar for qSH, confirming that qSH attenuation is modeled consistently by all kernels in the tested medium.

## Propagation raypaths and attenuation trajectories

The raypath models, which consist of overlaid trajectories from different kernels, provide a complementary overview of the numerical behavior of the three kernels at the global scale. These models are shown as the shortest paths (real raypaths) and maximum-energy damping trajectories (imaginary raypaths). For qP propagation raypaths (Fig. 11), the OEV and MEV are identical, as evidenced by completely overlapping trajectories, which is again consistent with their RMS misfits in $V^{Ray}$. However, the C-RRT rays exhibit subtle yet structured trajectory differences, which refract distinctly at all layer boundaries, and in some cases, they either overshoot or undershoot the receiver locations along flatter segments. Such deviations are concentrated at both the shallowest and deepest layer boundaries, with more obtuse critical angles than those from the other two kernels. Despite these differences, all raypaths correctly reproduce refraction events along the layer boundaries. The attenuation raypaths for qP further highlight the substantial similarity between MEV and OEV, as shown by overlapping paths through the domain (Fig. 12). They also illustrate that critical angles during refraction events are more acute for attenuation trajectories than for propagation trajectories at all layer boundaries. In our implementation, the propagation raypaths from C-RRT cannot be reconstructed because the corresponding traveltime field exhibits negative values (Fig. 6c).

For qSV propagation paths, the three kernels yield substantially different results (Fig. 13.a). The OEV captures the direct wave at near offsets but exhibits spurious reflections within the shallowest layer, occurring just above the layer boundary rather than along it. Consequently, reflections should not occur for the following two reasons. First, all nodes within the same layer are homogeneous, possessing the same dissipative anisotropic properties and ray quantities; thus, no property contrast exists to trigger a change in ray trajectory. Second, the current ray-tracing algorithm is designed to focus exclusively on first-arrival events. However, multi-arrival events could occur, although such solutions are not physically admissible in this configuration. Because each layer is homogeneous in our model with no parameter perturbation across the involved node pairs, true reflections cannot occur within a layer, and only refraction should occur at layer interfaces. The presence of reflections in the absence of an interface or mode-conversion source therefore indicates numerical artifacts rather than physical reflections, since reflection events are not physically expected. Thus, reflection should only occur when the reflection points lie along the boundary at near offsets, where the incident angles are still smaller than the critical angles. In our investigations, these 'pseudo-reflection' events can still be identified and even intensified, oscillating along the deeper layer boundaries at mid and far offsets, with negligible corresponding refraction events.

Meanwhile, the C-RRT produces propagation raypaths without pseudo-reflection and even includes a refraction event at the deepest layer boundary. However, it yields short, weakly directed refractions spanning only two to three nodes, with a westward direction that contradicts the wave propagation direction (Fig. 13.b). In contrast, the MEV solution showcases the most physically and numerically acceptable first-arrival qSV raypaths among others tested (Fig. 13.a). In qSV attenuation modeling, C-RRT raypaths cannot be generated because of the negative field values in the attenuation traveltime field. The OEV, however, produces attenuation paths that are broadly similar to those from MEV, as evidenced by overlapping refraction events and similar critical angles (Fig. 14a). However, this kernel still produces a pseudo-reflection just a few nodes above the shallowest layer boundary. Among the kernels tested, MEV therefore produces the most acceptable qSV trajectories.

Lastly, all kernels successfully model qSH propagation and attenuation along raypaths, with equivalent raypath locations in all qSH figures (Figs 15 and 16). Overall, these observations have several crucial implications. First, agreement in traveltime fields alone does not guarantee that the corresponding raypaths are physically correct. Kernel-level inconsistencies may still produce acceptable traveltime predictions while generating incorrect ray trajectories. This distinction is particularly important for applications such as traveltime tomography and ray-based migration, where inaccurate raypaths can lead to biased model updates even when traveltime residuals appear small. Second, these results demonstrate how significant kernel-level errors in C-RRT can distort the modeled medium geometry and lead to inaccurate accumulation of attenuation effects. Third, these findings emphasize the importance of MEV as the most stable kernel among the three examined, yielding the most physically acceptable and numerically stable ray quantities, raypaths, and traveltime.

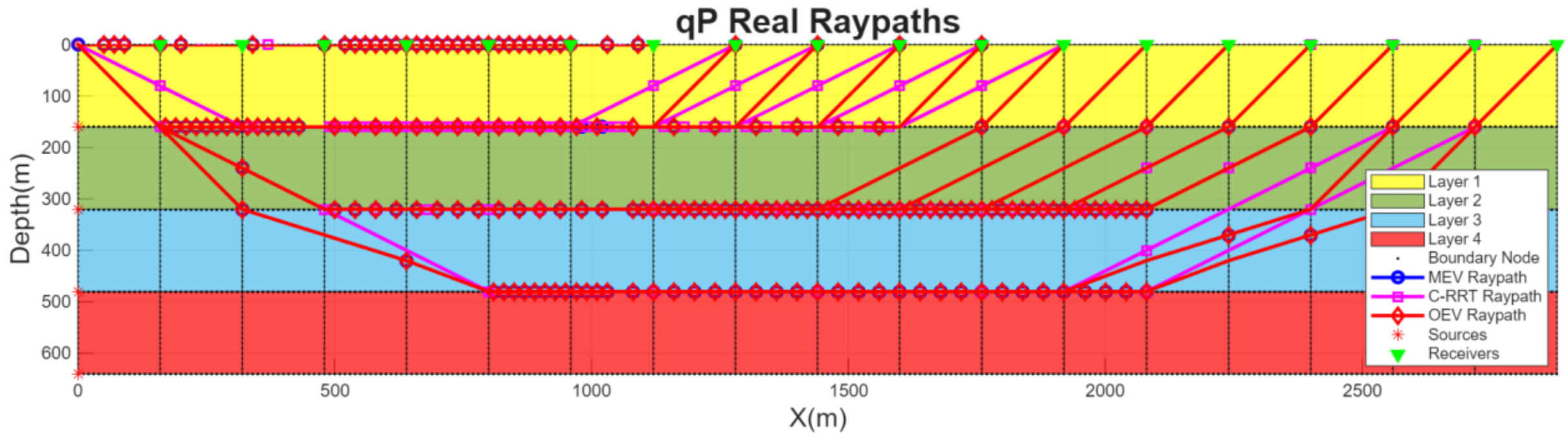

Fig. 11. qP propagation raypaths from OEV (red), C-RRT (magenta), and MEV (blue). OEV and MEV overlap almost everywhere, implying unperturbed field qP trajectories despite their small kernel-scale mismatches. C-RRT rays propagate differently across multiple interfaces and slightly perturb receiver-side trajectories, demonstrating how kernel-level errors alter field-scale geometry.

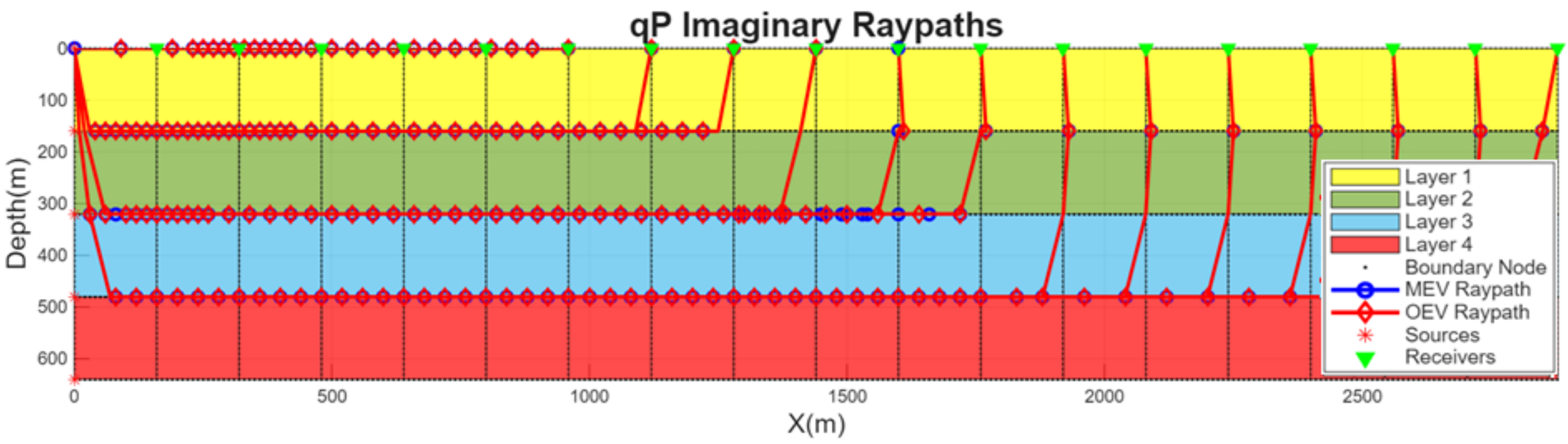

Fig. 12. qP attenuation raypaths from OEV (red) and MEV (blue). The two kernels yield overlapping attenuation paths and identical refraction behavior, while C-RRT raypaths are unable to be modelled because their negative qP attenuation is in traveltime units.

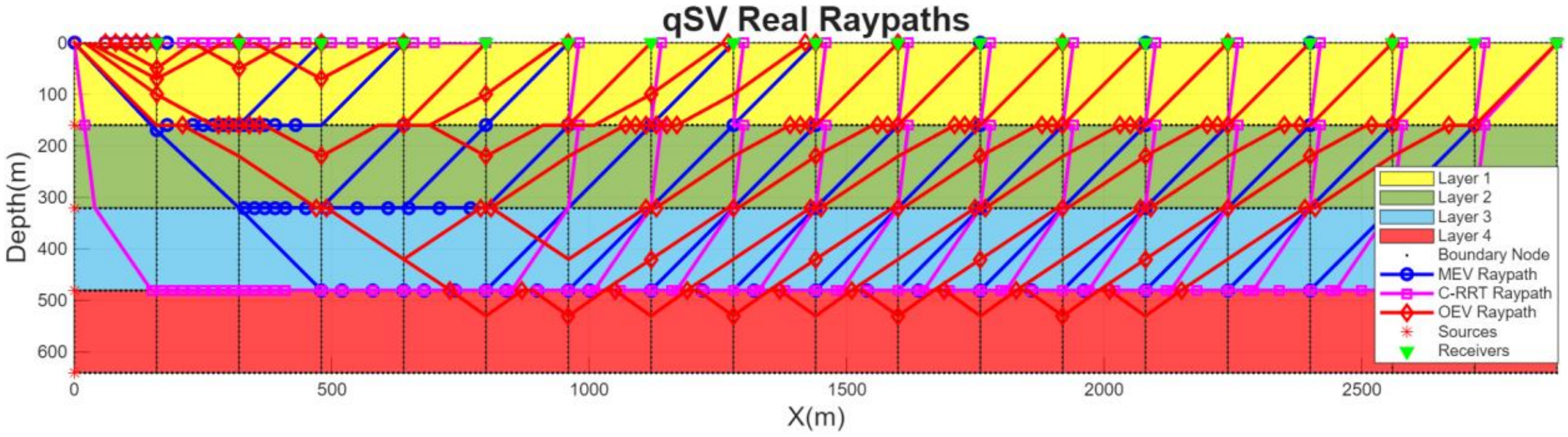

*a.*

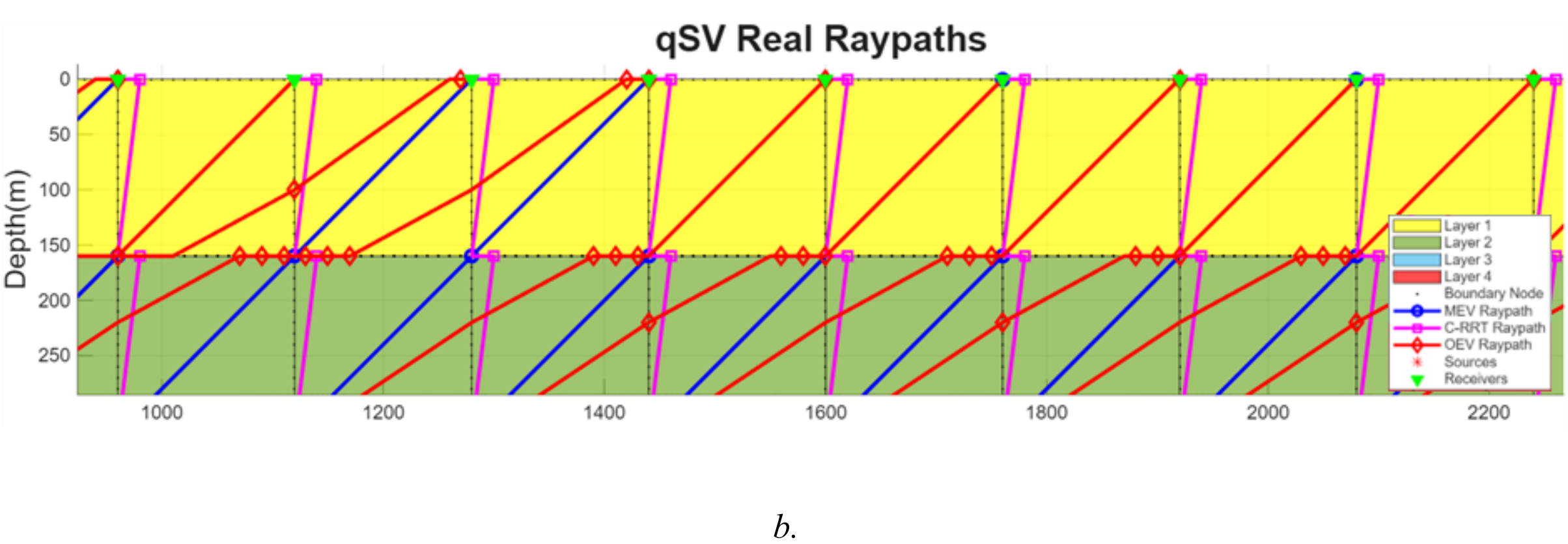

*b.*

Fig. 13. qSV propagation raypaths from all tested kernels: (a) full 2D model and (b) zoomed view. OEV (red diamond line) produces pseudo-reflections, C-RRT (magenta squarred-line) yields weak and partially inconsistent refractions, whereas MEV (blue circled line) provides the most geometrically and physically plausible first-arrival qSV raypaths.

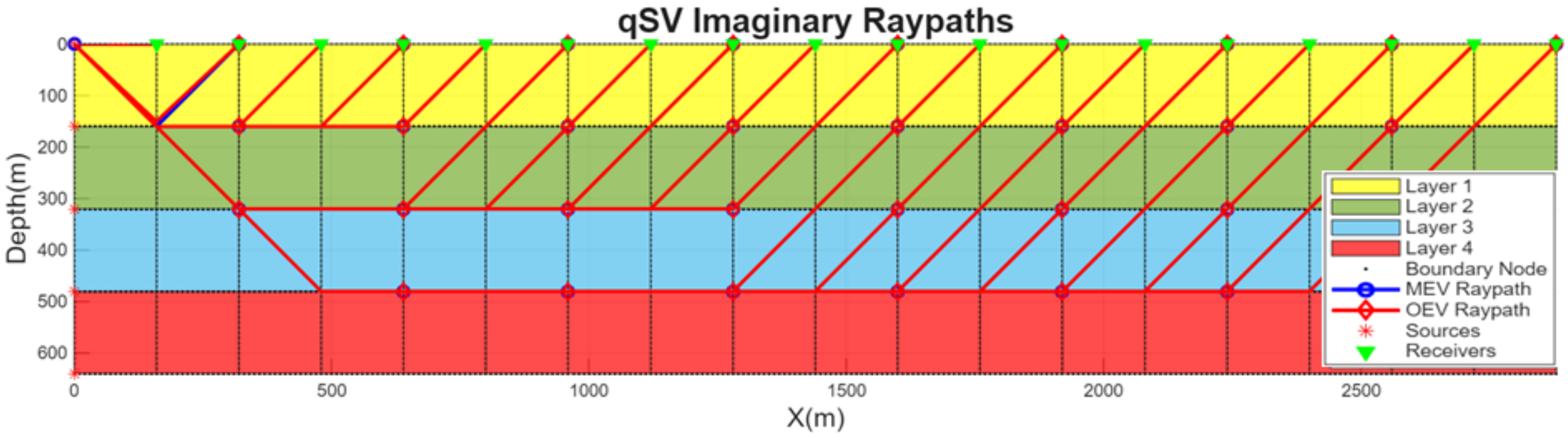

*a.*

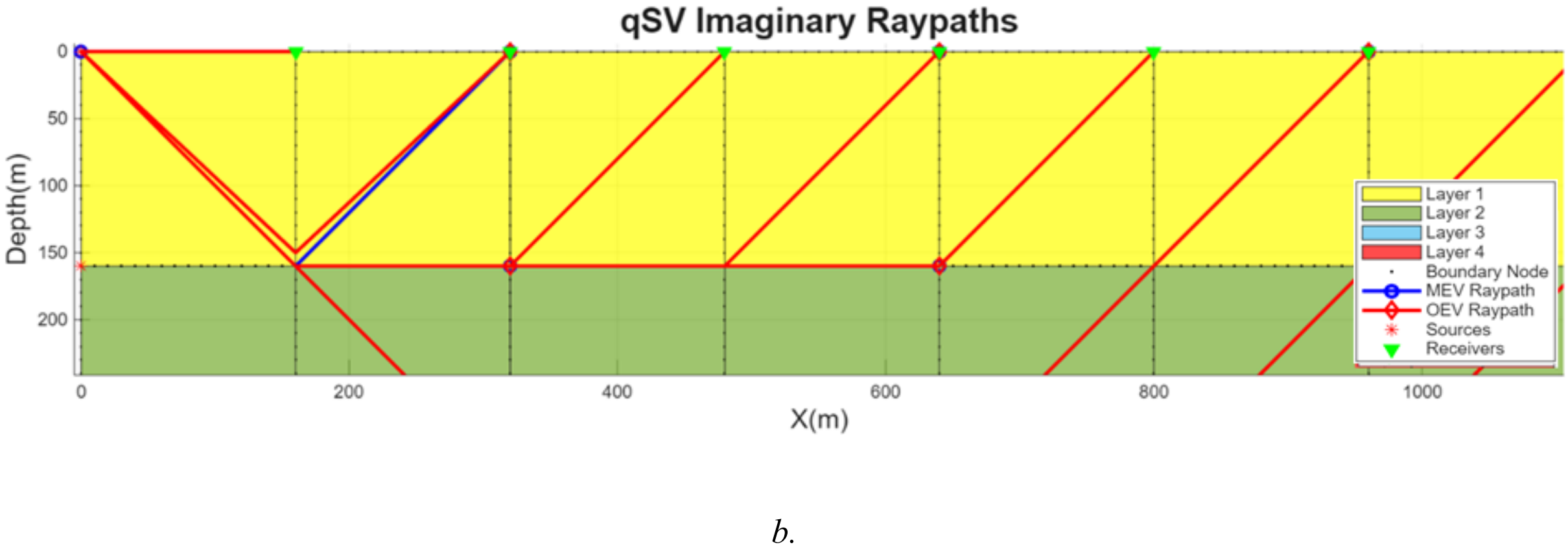

*b.*

Fig. 14. qSV attenuation raypaths: (a) 2D complete model and (b) zoomed view. MEV produces the most physically plausible attenuation paths among the tested kernels; OEV still shows pseudo-reflection artifacts, and C-RRT trajectories are absent because negative attenuation prevents possible raypath reconstruction.

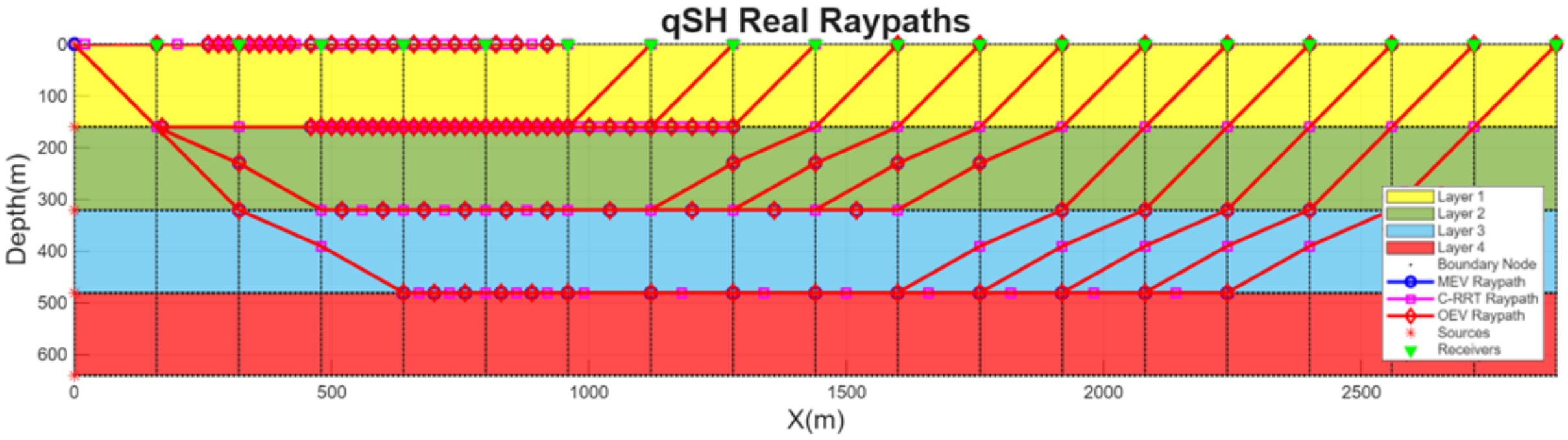

Fig. 15. qSH propagation raypaths from OEV (red), C-RRT (magenta), and MEV (blue). The near-perfect overlap demonstrates that all three kernels recover the same qSH first-arrival geometry.

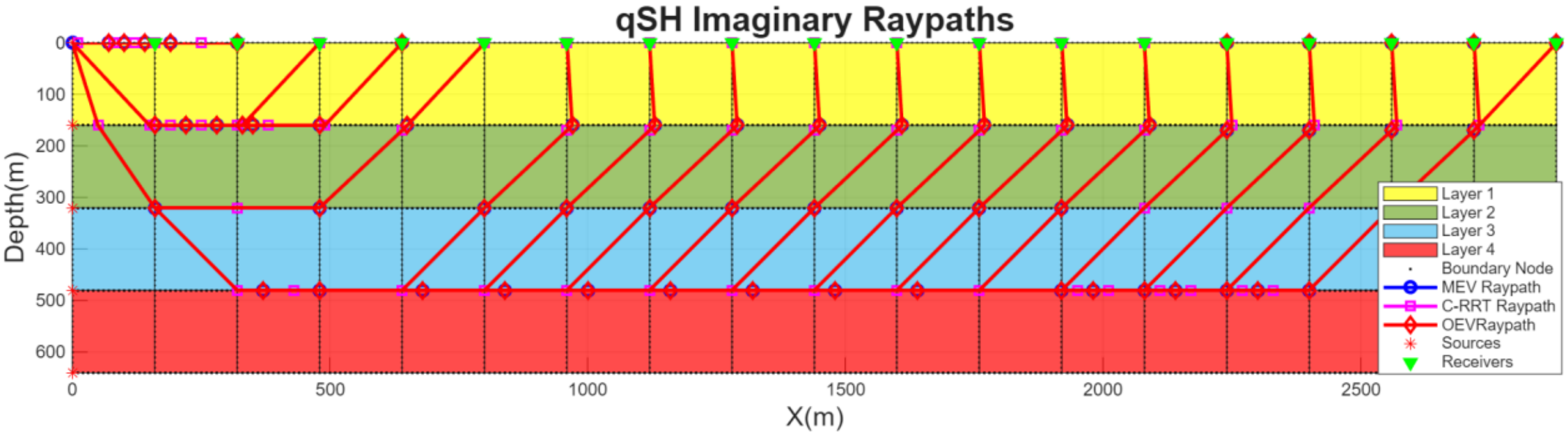

Fig. 16. qSH attenuation trajectories from OEV (red), C-RRT (magenta), and MEV (blue). The overlap of all trajectories confirms that qSH attenuation is stable across kernels and serves as a control case for the more challenging qP and qSV results.

## CONCLUSIONS

We have presented and evaluated a grid-based minimum-traveltime ray-tracing method for DAM, driven by a g*-Hamiltonian algorithm for the complex energy velocity. At the core of the framework is the modified energy-velocity kernel, obtained by enforcing the conditions of asymmetry and homogeneity on the complex density-normalized moduli and by computing the homogeneous complex energy velocity as a function of the complex slowness direction. This kernel function is then tailored to a Dijkstra-like shortest-path technique motivated by MSPM.

At the core level, we examined the performance of the OEV, C-RRT, and MEV kernels, using observed dissipative VTI rock properties. These numerical simulations show that the OEV and MEV produce comparable results for complex ray quantities, traveltime, and raypaths for qP and qSH. In contrast, the C-RRT kernel, although initially formulated to alleviate the qSV cusps and triplication challenges, can converge to spurious solutions of the slowness functional in the presence of attenuation and anisotropy. This results in significant errors in ray quantities and, in some cases, leads to non-physical negative attenuation in the traveltime unit. We also evaluated four optimization functions to reduce elapsed runtime, demonstrating that a one-dimensional bounded search provides a promising trade-off between robustness and computational efficiency and is therefore well-suited for large-scale tabulation of ray properties.

Tailoring these kernels into an MSPM-like shortest-path pipeline enables investigation of how local computational errors manifest at the field scale. Within the layered dissipative VTI model, for qP and qSH waves, the OEV- and MEV-driven propagation traveltime and attenuation in traveltime are almost identical, verifying that their small local misfits do not develop into significant field-scale biases. Their subsequent propagation raypaths and attenuation trajectories are also underpinned by their overlapping paths and correctly reproduce the expected refraction events at sharp velocity and density contrasts. In sharp contrast, C-RRT-based propagation traveltime contour maps for qP exhibit consistent patches of over- and under-shooting in regions of strong anisotropy and attenuation, and the qP attenuation in traveltime maps even indicates energy amplification rather than decay. These phenomena are strongly linked to specific angular ranges, with large RMS misfits relative to OEV, suggesting that maintaining the internal approximations of the C-RRT description is not computationally benign in the geological conditions of a DAM.

The qSV modeling, as the study's primary objective, provides the most rigorous validation. OEV-driven ray tracing tends to produce pseudo-reflections and other unrealistic paths in the presence of prominent qSV cusps. Meanwhile, the observed C-RRT implementation, despite having been proposed as a solution, fails to provide physically acceptable ray quantities and consistent complex slowness vector pairs at the kernel scale in our tests. It produces inconsistent attenuation models and induces non-physical refraction phenomena in the considered dissipative VTI model. In stark contrast, the MEV-based method produces the most physically acceptable and numerically stable qSV first-arrival events within the tested medium framework, ensuring that refracted rays occur at the correct interfaces, pseudo-reflections are suppressed, and attenuation remains consistent with the underlying medium parameters in heterogeneous DAM. These results confirm that, among the tested kernels, the full-model implementation of the g*-Hamiltonian(MEV)-driven approach provides the most stable modeling of all three-body waves in the considered DAM.

## DATA AND MATERIALS AVAILABILITY

Data associated with this research are available and can be obtained by contacting the corresponding author.